\documentclass[12pt,a4paper]{article}

\usepackage[a4paper,left=31mm,right=31mm,top=25mm,bottom=28mm]{geometry}
\usepackage{newtxtext,newtxmath}
\usepackage{amsmath,mathtools}
\usepackage{graphicx}
\usepackage{booktabs}
\usepackage{longtable}
\usepackage{array}
\usepackage{tabularx}
\usepackage{multirow}
\usepackage{caption}
\usepackage{enumitem}
\usepackage[numbers,sort&compress]{natbib}
\usepackage{microtype}
\usepackage{fancyhdr}
\usepackage{titlesec}
\usepackage{xcolor}
\usepackage[hidelinks]{hyperref}

\titleformat{\section}{\normalfont\bfseries\large}{\thesection.}{0.55em}{}
\renewenvironment{abstract}{\par\medskip\noindent\textbf{Abstract}\par\noindent\ignorespaces}{\par\medskip}

\newcommand{\Qn}{\{0,1\}^n}
\newcommand{\wt}{\operatorname{wt}}
\newcommand{\Ham}{d_{\mathrm H}}

\newcommand{\Gtf}{G_{\mathrm{TF}}}
\newcommand{\Gsec}{G_{\mathrm{sec}}}

\newcommand{\ket}[1]{|#1\rangle}

\newcommand{\topic}[1]{\par\medskip\noindent{\bfseries #1.}\par\noindent}
\newcommand{\Eqref}[1]{Eq.~\eqref{#1}}

\newcommand{\papertitle}{Sector-dominant graph-local drivers for path-window barrier Hamiltonians on the Boolean hypercube}
\newcommand{\paperauthors}{Takiko Sasaki$^{1,*}$ and Tetsuji Tokihiro$^{1}$}
\newcommand{\paperaffiliation}{$^{1}$Musashino University}
\newcommand{\paperemail}{t-sasaki@musashino-u.ac.jp; t-toki@musashino-u.ac.jp}

\newcommand{\makeioptitle}{%
\thispagestyle{plain}%
\begin{center}
{\bfseries\Large \papertitle\par}
\vspace{7mm}
{\large \paperauthors\par}
\vspace{3mm}
{\small \paperaffiliation\par}
\vspace{2mm}
{\small E-mail: \texttt{\paperemail}\par}
{\small Corresponding author: Takiko Sasaki, \texttt{t-sasaki@musashino-u.ac.jp}\par}
\vspace{4mm}
{\small arXiv public version; compiled \today\par}
\end{center}
\vspace{5mm}
}

\begin{document}
\makeioptitle

\begin{abstract}
We study finite-size adiabatic state preparation on Boolean hypercubes using graph-local drivers built from sector/path coordinates related to monotone Gray-code representatives.  The construction is not presented as a new all-$n$ Gray-code existence theorem; rather, it provides finite representatives, explicitly checked through the cases used in the numerical experiments, for testing problem-dependent graph-local drivers.  For ordinary diagonal-cost transverse-field annealing, the ordering does not yield a robust advantage, and we include this negative result as a baseline.  For non-diagonal target Hamiltonians whose geometry is expressed in the same sector/path coordinates, hybrid drivers combining sector, path-window, and small transverse-field components can substantially improve the final ground-state fidelity in centered barrier instances.  Reproduction runs from the accompanying code confirm a representative centered original-window barrier value of approximately \(0.9799\) for the fixed-control hybrid parameters \((w,\alpha,\epsilon)=(8,0.50,0.15)\), while also showing that the improvement is target-class dependent.  Randomized and ablation controls indicate that the dominant contribution is the sector-preserving skeleton, with strict one-bit completion acting as a secondary refinement.  We provide code, finite certificates, CSV files, validation logs, and reproduction scripts to make the finite-size claims traceable.
\end{abstract}


\section{Introduction}

Adiabatic quantum computation and quantum annealing use an interpolation
\begin{equation}
    H(s)=(1-s)H_D+sH_T,\qquad s=t/T,
\end{equation}
from an initial Hamiltonian \(H_D\) with an easily prepared ground state to a target Hamiltonian \(H_T\) whose ground state encodes the desired solution \cite{farhi2000,kadowaki1998,aharonov2007}.  The most common quantum-annealing driver for \(n\) binary variables is the transverse field
\begin{equation}
    H_{\mathrm{TF}}=-\sum_{i=1}^{n} X_i,
\end{equation}
where \(X_i\) is the Pauli-\(X\) operator acting on the \(i\)-th qubit.  In the computational basis it flips only the \(i\)-th bit.  Thus, from any bit string \(x\), the standard driver couples \(x\) to all \(n\) bit strings obtained by one-bit flips.  Equivalently, up to an additive constant, the transverse field is the graph Laplacian of the one-bit-flip hypercube.  More general problem-dependent drivers and catalyst terms have been studied as a way to change the geometry of the interpolation and, in some cases, improve finite-time behavior \cite{seki2012,hormozi2017}.

This paper investigates a sector/path coordinate on the Boolean hypercube \(\Qn\).  The strict ordering, denoted \(E^{\mathrm{orig}}_n\), arranges bit strings according to a layered Hamming-weight skeleton and a fixed initial strip.  Its one-bit adjacency and rank-by-rank progression through Hamming-weight levels put it in the class of monotone Gray codes.  Savage and Winkler introduced monotone Gray codes as Hamilton paths in the Boolean lattice whose edges between levels \(i-1\) and \(i\) appear before edges between levels \(i\) and \(i+1\), and proved the existence of such orderings for all \(n\) \cite{savageWinkler1995}.  Therefore the present work is not a new all-\(n\) Gray-code existence theorem.  Instead, we use a particular fixed-prefix, fixed-skeleton representative of this known class.  For \(n\le 8\), the representative is obtained by deterministic depth-first completion and the \(n=8\) strict path is checked by the compiled generator log.  A corresponding strict \(n=9\) completion was attempted but did not complete within the attempted computation, so \(n=9\) is not used as a completed strict-generator theorem.  A fast greedy variant, denoted \(E^{\mathrm{v2}}_n\), preserves the same sector skeleton and fixed prefix for general \(n\), but it no longer enforces one-bit adjacency at every step.

The point of formalizing the ordering is explicit.  We want a finite-size construction that simultaneously exposes the Hamming-weight sector \(\wt(x)\), provides a discrete path coordinate \(p_E(x)\), and can therefore be used both to define graph-local adiabatic drivers and to band sparse Hamiltonians whose off-diagonal structure is local in that coordinate.  In other words, the object of interest is not merely the list of bit strings itself, but the geometry induced by sector plus path information.

The first possible interpretation of such an ordering is as a better labeling of classical configurations.  If neighboring integer labels are mapped to one-bit-neighboring bit strings, then one might hope that ordinary transverse-field annealing follows the intended integer path.  Our simulations show that this interpretation is not correct in general.  The transverse-field driver does not know the integer label.  It moves on the full hypercube, and a chosen Hamiltonian path is only a small subgraph of that hypercube.  Consequently, smoothness of the encoding is insufficient to guarantee better diagonal-cost annealing.

The second interpretation, which is the one pursued here, is to use the ordering as a geometric object.  It provides a path coordinate \(p_E(x)\) and therefore a family of path-window graphs on \(\Qn\).  Combining this path geometry with a graph that respects Hamming-weight sectors gives a problem-dependent graph-local driver.  The main conclusion is that a sector-dominant hybrid driver can improve finite-time ground-state preparation for non-diagonal targets whose locality is aligned with a path-window barrier structure.  The improvement is conditional: path-only drivers are weak, several target classes or geometries can favor the transverse-field or sector-only baselines, and the existence of monotone Gray codes does not by itself imply that the particular driver geometry studied here is useful.  The conclusion supported by the data is therefore a graph-local driver design principle, not a universal encoding theorem.

We do not claim a new infinite Gray-code family, because the relevant monotone Gray-code existence theory is already known.  We also do not claim a universal relabeling for ordinary diagonal transverse-field annealing, or a hardware-ready driver for present-day transverse-field Ising machines.  The sector and path-window Laplacians used here are custom graph-local Hamiltonians on configuration space.  The paper should therefore be read as a finite-size numerical design study for structured non-diagonal Hamiltonians rather than as a universal annealing-improvement theorem.

\section{Sector-snake coordinate as a monotone Gray-code representative}

Let
\begin{equation}
    Q_n=\{0,1\}^n
\end{equation}
be the \(n\)-bit state space.  For each bit string \(x\in Q_n\), we write \(\ket{x}\) for the corresponding computational-basis vector of the \(2^n\)-dimensional Hilbert space.  We identify a bit string \(x\in Q_n\) with a subset \(S_x\subseteq\{1,\ldots,n\}\), where \(i\in S_x\) if and only if \(x_i=1\).  Its Hamming weight is
\begin{equation}
    \wt(x)=|S_x|=\sum_{i=1}^n x_i .
\end{equation}
The Hamming-weight sector of weight \(j\) is
\begin{equation}
    V_j=\{x\in Q_n:\wt(x)=j\},\qquad |V_j|=\binom{n}{j}.
\end{equation}
The notation \([j:k]\) means the \(k\)-th block or position inside the sector \(V_j\).  The point of this notation is that a bit string has two pieces of coarse information: the number \(j\) of selected elements and its relative position inside the sector.  Problems with cardinality constraints, support-size constraints, or staged addition/removal of elements often have this type of structure.

An ordering is a bijection
\begin{equation}
    E_n:\{0,1,\ldots,2^n-1\}\to Q_n.
\end{equation}
Its inverse path coordinate is written as
\begin{equation}
    p_E(x)=E_n^{-1}(x).
\end{equation}
When a matrix is indexed by bit strings, the function \(p_E\) is also the row and column order used to display or truncate the matrix.  When a graph is defined by path distance, \(p_E\) becomes a reaction coordinate on the discrete state space.

We use the following \emph{sector-snake coordinate}.  At the level of one-bit adjacency and monotone progression through Hamming-weight sectors, the strict sector-snake ordering is a monotone Gray code.  More precisely, Savage and Winkler define a monotone Gray code as an enumeration of all \(n\)-bit strings by one-bit moves such that edges between adjacent levels occur in level order; equivalently, edges between levels \(i-1\) and \(i\) appear before edges between levels \(i\) and \(i+1\) \cite{savageWinkler1995}.  The strict sector-snake skeleton used below has exactly this rank-by-rank form: it alternates between levels \(k\) and \(k+1\) during stage \(k\), and only then proceeds to the next pair of levels.  Thus the strict sector-snake ordering should be regarded as a fixed-prefix, fixed-skeleton representative of the known monotone Gray-code class, not as a new Gray-code existence theorem.

This distinction matters for the novelty claim.  The all-\(n\) existence of some monotone Gray code is known from the work of Savage and Winkler.  What is specific to the present paper is different: we select concrete finite representatives by deterministic or certificate-based completion rules; we impose a particular fixed prefix and active-count skeleton; and, most importantly, we use the resulting sector/path coordinate to define path-window Hamiltonians, graph-local drivers, and banded matrix representations.  The binary reflected Gray code is a Hamiltonian path or cycle of the hypercube, but it does not expose the same binomial Hamming-weight skeleton.  Constant-weight combination Gray codes, including revolving-door and cool-lex type orders, list a single sector \(V_j\) or a related fixed-content family, whereas the present coordinate interleaves all Hamming-weight sectors into a single monotone path \cite{ruskey2009,mutze2023}.  The broader literature on Boolean-lattice Hamiltonicity, symmetric chain decompositions, and middle-level or weight-interval Gray codes is closely related and provides the correct combinatorial background \cite{savage1997,streib2014,gregor2018,mutze2023}.

The purpose of formalizing the coordinate is explicit.  We want an ordering that simultaneously exposes the Hamming-weight sector \(\wt(x)\), yields a one-bit monotone path, supplies a discrete path coordinate \(p_E(x)\) for defining path-window graphs and barrier landscapes, and bands matrices whose off-diagonal couplings are local in that coordinate.  The numerical sections below test these intended uses directly.

The strict sector-snake representative is built from a sector skeleton
\begin{equation}
    \pi^{(n)}=(j_0,j_1,\ldots,j_{2^n-1}),
\end{equation}
where \(j_t=\wt(E_n(t))\).  A convenient equivalent description uses the \emph{active counts}
\begin{equation}
    a_0=1,\qquad a_k=\binom{n}{k}-a_{k-1}+1,\qquad k=1,\ldots,n.
\end{equation}
At stage \(k\), the skeleton places \(a_k\) active \(k\)-subsets and \(a_k-1\) bridge \((k+1)\)-subsets in the alternating pattern
\begin{equation}
    k,(k+1),k,(k+1),\ldots,(k+1),k.
\end{equation}
The alternating-stage description is used for \(k=0,\ldots,n-1\), omitting empty terminal stages; if \(a_n=1\), the final \(n\)-subset appears as the terminal active state rather than as the start of a nonexistent \((n+1)\)-bridge stage.  Because
\begin{equation}
    a_k+(a_{k-1}-1)=\binom{n}{k},
\end{equation}
each Hamming-weight sector appears exactly the required number of times.  Thus the skeleton is a prescribed sector-snake interleaving of neighboring Hamming-weight sectors.  The path starts with the fixed prefix
\begin{equation}
\emptyset,\ \{1\},\ \{1,2\},\ \{2\},\ \{2,3\},\ldots,\ \{n-1,n\},\ \{n\}.
\end{equation}
The remaining states are filled so as to satisfy
\begin{equation}
    \wt(E_n(t))=j_t,\qquad
    \Ham(E_n(t),E_n(t+1))=1,
\end{equation}
and bijectivity over all \(2^n\) states.  We call this the strict or original representative.  The word ``representative'' is used deliberately: the monotone Gray-code class exists independently of the present search rule, whereas the present computations use one concrete fixed-prefix representative of that class.

\topic{Rank-by-rank bridge formulation}
The sector skeleton has a useful rank-by-rank interpretation.  At stage \(k\), the active \(k\)-subsets can be viewed as vertices of the Johnson graph \(J(n,k)\), whose vertices are \(k\)-subsets of \(\{1,\ldots,n\}\) and whose edges connect two subsets that differ by the exchange of one element.  If consecutive active states are \(X\) and \(X'\) with \(|X|=|X'|=k\) and \(|X\cap X'|=k-1\), then
\begin{equation}
    Y=X\cup X'
\end{equation}
is a \((k+1)\)-subset and gives the two-step hypercube move
\begin{equation}
    X\to Y\to X'.
\end{equation}
Thus each Johnson edge \((X,X')\) carries the bridge color \(Y=X\cup X'\).  A strict stage can therefore be viewed as a Hamilton path through the active \(k\)-subsets whose edge colors are all distinct and unused, together with endpoint constraints inherited from the neighboring ranks.  In this language, the finite completion problem becomes a sequence of edge-colored rainbow Hamilton-path problems on Johnson graphs, coupled across ranks by the fixed endpoints and bridge counts.

\topic{Deterministic completion rule for the canonical \(n\le 8\) paths}
For reproducibility, we state the completion rule explicitly.  Suppose that a partial path \(x_0,\ldots,x_{t-1}\) has already been built.  The candidate set for the next state is
\begin{equation}
    \mathcal C_t=\{y\notin\{x_0,\ldots,x_{t-1}\}:\wt(y)=j_t,\ \Ham(y,x_{t-1})=1\}.
\end{equation}
If \(t<2^n-1\), each candidate \(y\in\mathcal C_t\) is given an onward-count score
\begin{equation}
    c_t(y)=\#\{z\notin\{x_0,\ldots,x_{t-1},y\}:\wt(z)=j_{t+1},\ \Ham(z,y)=1\}.
\end{equation}
Let \(y=x_{t-1}\triangle\{i\}\), where \(i\) is the unique flipped bit.  The tie-breaking key is
\begin{equation}
    \tau_t(y)=
    \begin{cases}
        -i, & \wt(y)=\wt(x_{t-1})+1 \quad \text{(prefer larger added index)},\\
        \ \ i, & \wt(y)=\wt(x_{t-1})-1 \quad \text{(prefer smaller removed index)}.
    \end{cases}
\end{equation}
Candidates are explored in increasing lexicographic order of \((c_t(y),\tau_t(y),y)\).  The first candidate leading to a full path of length \(2^n\) is accepted; if no candidate succeeds, the search backtracks.  This deterministic rule produces the canonical strict paths used in the \(n=5,6,7,8\) tables and numerical experiments.

\topic{Finite-size strict-generator validation}
Although the existence of monotone Gray codes is known for all \(n\), the all-\(n\) success of our specific fixed-prefix completion rule is not proved.  The reproducible strict representative used in the main numerical scans is explicitly validated through \(n=8\).  In the compiled check, the \(n=8\) strict generator completed a valid path of length \(256\), with fixed prefix, skeleton, bijectivity, and one-bit adjacency verified.  The selected deterministic representative is not claimed to have a completed strict \(n=9\) certificate; the \(n=9\) attempt and timeout log are included in the reproduction package.  Thus the \(n=9\) material in the archive is a finite attempted computation and validation log, not a completed strict-generator certificate and not an asymptotic scaling claim.

The fast \(v2\) ordering preserves the same skeleton, fixed prefix, and bijectivity, but replaces exact depth-first completion by a greedy rule.  If \(U_{j_t}\) is the set of unused states of the required weight \(j_t\), then \(v2\) chooses
\begin{equation}
    x_t=\arg\min_{y\in U_{j_t}} \bigl(\Ham(y,x_{t-1}),\operatorname{span}(y),y\bigr),
\end{equation}
where \(\operatorname{span}(y)\) is the difference between the largest and smallest selected indices in \(S_y\) for nonempty \(y\), and zero for the empty set.  This gives a scalable practical completion but introduces occasional jumps of Hamming distance three for the sizes studied.  For this reason, \(v2\) should not be described as the same construction as the strict ordering.  It is a useful control and a possible scalable variant, but it relaxes the one-bit path constraint.

\begin{figure}[htbp]
\centering
\includegraphics[width=.49\linewidth]{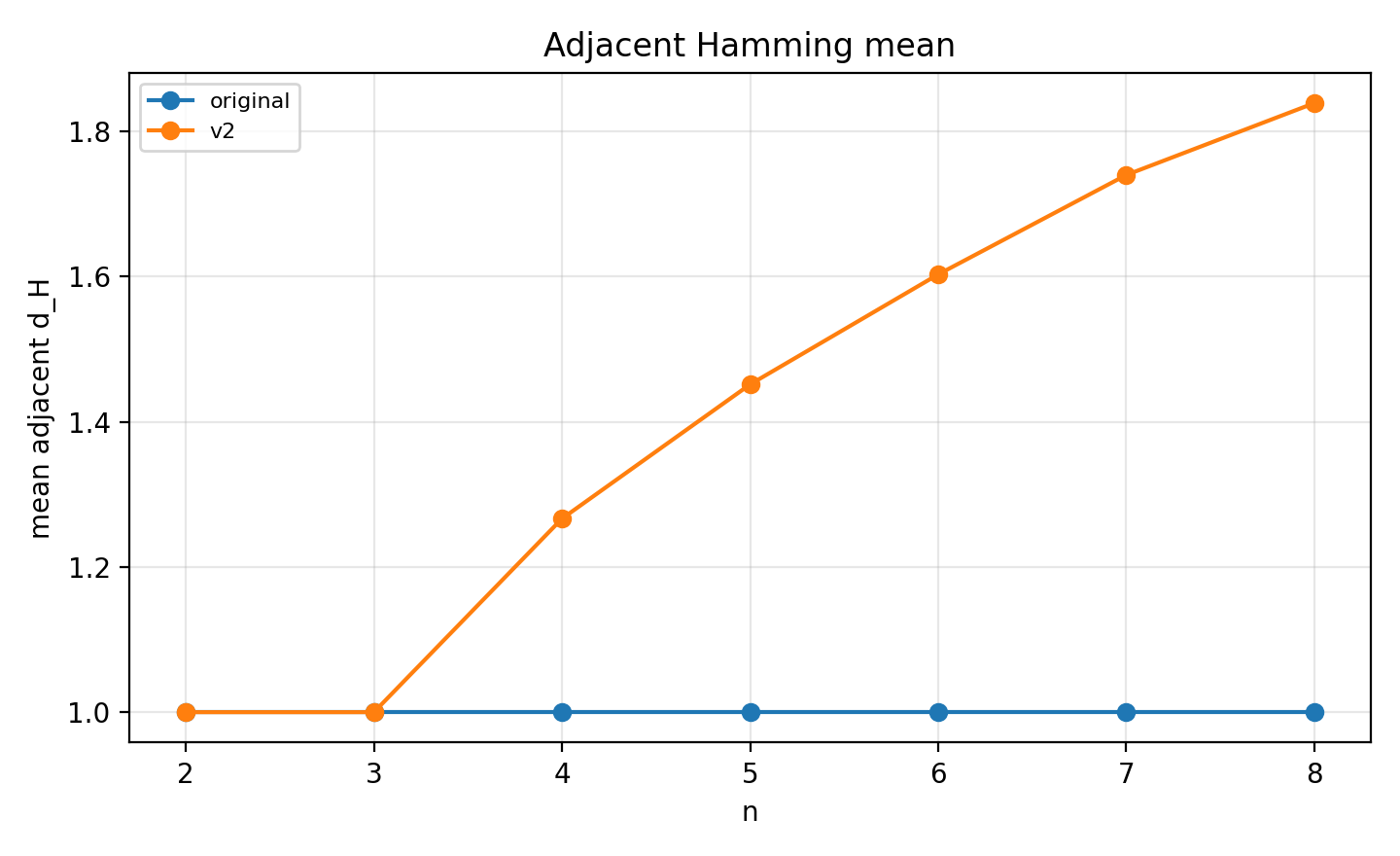}\hfill
\includegraphics[width=.49\linewidth]{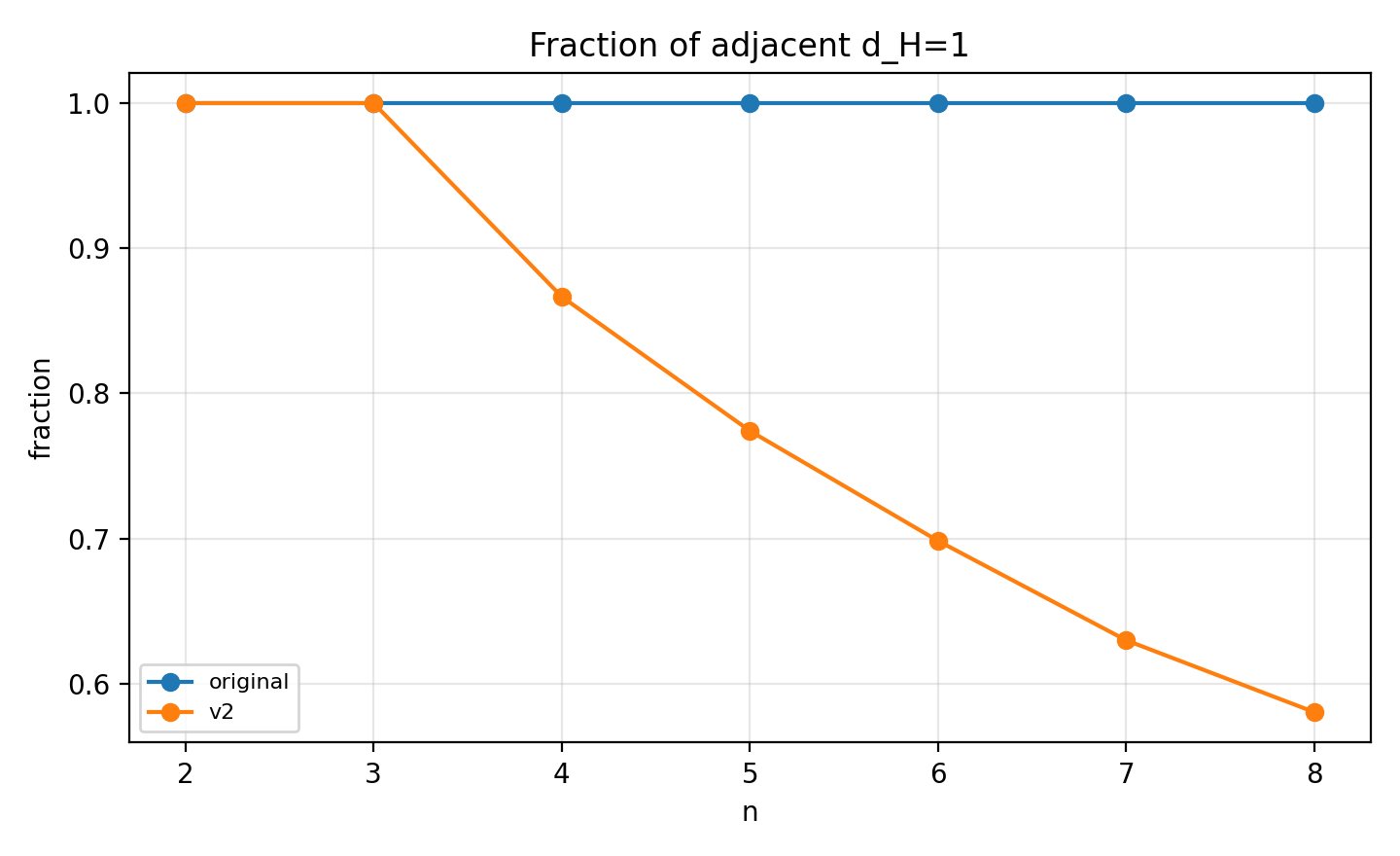}
\caption{Locality diagnostics for the strict sector-snake ordering and the fast \(v2\) ordering in the regenerated \(n\le 8\) benchmark.  The strict ordering keeps all adjacent moves at Hamming distance one in these checked cases.  The \(v2\) ordering preserves the skeleton and fixed prefix but introduces longer adjacent jumps.}
\label{fig:generator}
\end{figure}

\begin{table}[htbp]
\centering
\caption{Generator diagnostics for the strict ordering and the fast \(v2\) ordering in the regenerated \(n\le 8\) benchmark.}
\label{tab:generator}
\begin{tabular}{ccccc}
\toprule
\(n\) & ordering & mean adjacent \(d_H\) & max adjacent \(d_H\) & fraction \(d_H=1\)\\
\midrule
5 & strict & 1.000 & 1 & 1.000\\
5 & \(v2\) & 1.452 & 3 & 0.774\\
6 & strict & 1.000 & 1 & 1.000\\
6 & \(v2\) & 1.603 & 3 & 0.698\\
7 & strict & 1.000 & 1 & 1.000\\
7 & \(v2\) & 1.740 & 3 & 0.630\\
8 & strict & 1.000 & 1 & 1.000\\
8 & \(v2\) & 1.839 & 3 & 0.580\\
\bottomrule
\end{tabular}
\end{table}

\begin{table}[htbp]
\centering
\caption{Strict-generator validation status in the regenerated reproduction package.}
\label{tab:strict-validation}
\begin{tabular}{>{\raggedright\arraybackslash}p{0.18\linewidth}>{\raggedright\arraybackslash}p{0.30\linewidth}>{\raggedright\arraybackslash}p{0.42\linewidth}}
\toprule
case & status & log summary\\
\midrule
\(n=8\) & completed strict path & path length \(256\), validation OK, compiled generator completed with \(65{,}717\) searched nodes\\
\(n=9\) & attempted but not completed & timeout/status log included in the reproduction package; no completed strict path or certificate is claimed\\
\bottomrule
\end{tabular}
\end{table}

Tables~\ref{tab:generator} and~\ref{tab:strict-validation} clarify the finite-size status of the construction.  The strict ordering is explicitly checked through the cases used in the reproduced numerical scans, with the strict \(n=8\) path validated by the compiled generator log.  The strict \(n=9\) attempt is recorded only as a timeout/status log in the reproduction package.  The fast ordering is less local as a path, but it remains useful for testing whether the effect is simply due to using any sector skeleton or whether the strict path itself matters.

\section{Graph-local Hamiltonians}

For a simple graph \(G=(Q_n,E)\), let \(A(G)\) be its adjacency matrix and \(D(G)\) its diagonal degree matrix.  We use the graph Laplacian, a standard object in spectral graph theory \cite{chung1997},
\begin{equation}
    L(G)=D(G)-A(G).
\end{equation}
In the simulations, each graph Laplacian is scaled to unit spectral norm,
\begin{equation}
    \widehat L(G)=\frac{L(G)}{\lambda_{\max}(L(G))},
\end{equation}
where \(\lambda_{\max}(L(G))\) denotes the largest eigenvalue of the positive-semidefinite Laplacian \(L(G)\).  This normalization is not essential to the definition, but it is useful numerically because the dense sector graph and the sparse path-window graph have very different degrees.  With this scaling, the coefficients \(\alpha\) and \(\epsilon\) in the hybrid driver can be interpreted as relative mixture weights rather than as artifacts of graph size.

If \(G\) is connected, the uniform superposition over computational basis states
\begin{equation}
    |u\rangle=2^{-n/2}\sum_{x\in Q_n}|x\rangle
\end{equation}
is the zero-eigenvalue ground state of \(L(G)\).  Here each \(\ket{x}\) is the computational-basis state indexed by the bit string \(x\in Q_n\).  This is the reason graph Laplacians are convenient driver Hamiltonians: the initial state is simple even when the allowed transitions are not the standard one-bit transitions.

We use three basic graphs.  First, the transverse-field graph \(\Gtf\) is the hypercube graph,
\begin{equation}
    (x,y)\in E(\Gtf)\quad\Longleftrightarrow\quad \Ham(x,y)=1.
\end{equation}
Its Laplacian satisfies
\begin{equation}
    L(\Gtf)=nI-\sum_{i=1}^{n}X_i,
\end{equation}
so it is equivalent to the standard transverse-field driver up to an additive constant and scaling.  Second, the sector graph \(\Gsec\) connects all states whose Hamming weights differ by at most one:
\begin{equation}
    (x,y)\in E(\Gsec)\quad\Longleftrightarrow\quad x\ne y,\quad |\wt(x)-\wt(y)|\le 1 .
\end{equation}
This graph is dense.  It is not intended as a direct hardware-local driver; rather, it models broad transport among configurations with nearby cardinalities.  Third, given an ordering \(E\), the path-window graph \(G_{E,w}\) connects states close in the path coordinate:
\begin{equation}
    (x,y)\in E(G_{E,w})\quad\Longleftrightarrow\quad
    0<|p_E(x)-p_E(y)|\le w,\quad |\wt(x)-\wt(y)|\le 1 .
\end{equation}
The final sector condition matches the sector-interleaving construction and removes long jumps across distant weight layers.

The main driver is
\begin{equation}
\label{eq:hybrid-driver}
    H_D(\alpha,\epsilon,w)
    =
    (1-\epsilon)\left\{(1-\alpha)\widehat L(\Gsec)
    +\alpha\widehat L(G_{E,w})\right\}
    +\epsilon \widehat L(\Gtf).
\end{equation}
Here \(w\) is the driver-window width, \(\alpha\in[0,1]\) weights the path-window component, and \(\epsilon\in[0,1]\) weights the transverse-field catalyst.  The sector graph is the dominant transport graph.  It prevents the dynamics from being confined to a thin one-dimensional chain.  The path-window component is a catalyst: it gives the driver some knowledge of the reaction coordinate or staged path used by the target Hamiltonian.  The transverse-field term is kept small and acts as a conventional one-bit-flip catalyst.  This interpretation is important because the numerical results below show that the path-window graph alone is a poor driver.

The main target family is a non-diagonal path-window barrier Hamiltonian
\begin{equation}
\label{eq:target}
    H_T=\widehat L(G_{E,w_T})+V_{p_*,h}.
\end{equation}
For \(N=2^n\), \(V_{p_*,h}\) is diagonal in the computational basis and depends on \(p_E(x)\).  Before the final normalization of the potential range, it has the form
\begin{equation}
    V_{p_*,h}(x)
    =
    \frac{|p_E(x)-p_*|}{N-1}
    +
    h\exp\left[-\left(\frac{p_E(x)-p_b}{\sigma N}\right)^2\right],
\end{equation}
with \(p_b=0.35(N-1)\) and \(\sigma=0.06\).  The parameter \(p_*/(N-1)\) specifies the target position on the path and \(h\) controls the barrier height.  The barrier-location parameter \(p_b\) is chosen empirically so that, for the centered benchmark \(p_*/(N-1)=0.50\), the barrier sits distinctly away from the target minimum and from the endpoints of the path, thereby creating a nontrivial intermediate obstruction rather than an endpoint artifact.  The width parameter \(\sigma\) is chosen to make the Gaussian barrier broad on the scale of a single path vertex but still narrow compared with the full path length; numerically, \(\sigma=0.06\) gives a barrier spanning only a modest fraction of the \(2^n\) states.  These values are fixed benchmark choices carried over through the scans; they are not themselves optimized parameters.  The potential is shifted and rescaled so that its minimum is zero and its maximum is one.  This target is non-diagonal because the Laplacian term already contains off-diagonal transitions.  It should be viewed as a ground-state-preparation Hamiltonian on a structured graph, not as an ordinary diagonal QUBO.

\section{Numerical method}

We simulate the time-dependent Schroedinger equation with a linear schedule
\begin{equation}
    H(s)=(1-s)H_D+sH_T,\qquad s=t/T.
\end{equation}
The initial state is the uniform ground state of \(H_D\), which is available because all driver components are graph Laplacians on connected graphs in the cases considered.  Time evolution is computed by piecewise-constant midpoint propagation using matrix exponential actions.  Sparse representations are used for the sparse components, and at the present finite sizes direct dense routines are also used whenever convenient.  For the main \(n=8\) scans, we use \(T=80\) and 35 midpoint slices.  Ground states and low gaps are computed either by dense diagonalization for small matrices or by sparse eigensolvers.

The principal metric is the target ground-state fidelity
\begin{equation}
    F=|\langle \phi_0|\psi(T)\rangle|^2,
\end{equation}
where \(\phi_0\) is the normalized ground state of \(H_T\).  We also report the energy residual
\begin{equation}
    \langle\psi(T)|H_T|\psi(T)\rangle-E_0.
\end{equation}
Fidelity is the cleanest measure when the purpose is ground-state preparation.  The residual is a complementary measure because two states with similar fidelity can still differ in their final target energy if the target spectrum is dense.

The numerical study is finite-size.  The detailed parameter scans reported below are centered at \(n=8\), so the largest Hilbert space used in the main regenerated scan study has dimension \(2^8=256\).  The strict \(n=8\) path is validated by the compiled generator log.  The strict \(n=9\) attempt did not complete within the attempted computation, and no asymptotic scaling claim is made from it.

As a basic convergence check, table~\ref{tab:convergence} reports the centered strict-window barrier benchmark at \(n=8\), \(T=80\) for several numbers of midpoint slices.  The reported fidelities are stable to the fourth decimal place for the main drivers, so the differences discussed later are not artifacts of the time discretization.

\begin{table}[htbp]
\centering
\caption{Midpoint-slice convergence check on the centered strict-window barrier target at \(n=8\), \(T=80\).}
\label{tab:convergence}
\begin{tabular}{lccc}
\toprule
driver & 35 slices & 70 slices & 140 slices\\
\midrule
TF only & 0.8902 & 0.8901 & 0.8901\\
sector only & 0.9455 & 0.9453 & 0.9453\\
sector + path, no TF & 0.9697 & 0.9695 & 0.9694\\
sector + path + TF & 0.9799 & 0.9797 & 0.9797\\
path only & 0.2490 & 0.2489 & 0.2488\\
\bottomrule
\end{tabular}
\end{table}

\section{Standard transverse-field diagonal QA is not improved by the ordering}

The first test used ordinary diagonal costs, as in standard Ising/QUBO-style formulations of combinatorial optimization \cite{lucas2014}.  In this setting
\begin{equation}
    H_T=\operatorname{diag}(C_E),\qquad
    H_D=-\sum_i X_i,
\end{equation}
where \(C_E\) is the same logical cost arranged in the computational basis according to an encoding \(E\).  Encodings compared included binary order, Gray order, the strict sector-snake order, and \(v2\).

The four diagonal-cost families in table~\ref{tab:negative-diagonal} are defined on the logical path coordinate \(t\in\{0,\ldots,N-1\}\), where \(N=2^n\).  Let \(t_*\) be the selected target index and let \(\pi_t\) denote the Hamming-weight sector assigned to the \(t\)-th path position.  We set
\[
 d_{\rm idx}(t)=\frac{|t-t_*|}{N-1},\qquad
 d_{\rm sec}(t)=\frac{|\pi_t-\pi_{t_*}|}{n}.
\]
The raw costs are
\[
 C_{\rm idx}(t)=d_{\rm idx}(t),\qquad
 C_{\rm sec}(t)=d_{\rm sec}(t)+0.02\,d_{\rm idx}(t),
\]
\[
 C_{\rm mix}(t)=\frac12 d_{\rm sec}(t)+\frac12 d_{\rm idx}(t),
\]
and
\[
 C_{\rm bar}(t)=d_{\rm idx}(t)+0.35
 \exp\!\left[-\left(\frac{t-(N-1)/2}{0.10N}\right)^2\right],
 \qquad C_{\rm bar}(t_*)=0.
\]
Each raw cost is shifted and rescaled to \([0,1]\) before being assigned to computational bit strings through the chosen encoding.  These definitions match the reproduction script \texttt{02\_qa\_simulator\_original\_v2\_n8.py}.

The result was not a consistent advantage for either sector/path ordering.  This negative result is conceptually important.  A chosen ordering gives a path through the hypercube, while the transverse-field driver uses the entire hypercube.  Each state has \(n\) transverse-field neighbors, only one or a few of which are adjacent along the selected ordering path.  Therefore, even if
\[
    \Ham(E(t),E(t+1))=1
\]
for all \(t\), the standard driver is not forced to move from \(E(t)\) to \(E(t+1)\).  It can move in any one-bit direction.  Consequently, the ordering is not a universal improvement for standard diagonal-cost quantum annealing.  This is the reason the rest of the paper studies non-diagonal targets and custom graph-local drivers.

Table~\ref{tab:negative-diagonal} makes this negative result quantitative for \(n=8\), \(T=80\).  Each entry is the mean target-success probability over the three target positions \(p_*/(N-1)=0.25,0.50,0.75\) for the indicated diagonal-cost family.  The strict and \(v2\) orderings are not robustly better than binary or Gray order, and in several families they are clearly worse.  These numbers should be read as reference evidence for what the proposed coordinate does \emph{not} do.

\begin{table}[htbp]
\centering
\caption{Standard transverse-field diagonal QA at \(n=8\), \(T=80\): mean target-success probability over three target positions for several diagonal-cost families.  Larger values are better.}
\label{tab:negative-diagonal}
\begin{tabular}{lcccc}
\toprule
cost family & binary & Gray & strict & \(v2\)\\
\midrule
index well & 0.0252 & 0.0247 & 0.0176 & 0.0211\\
sector well & 0.0135 & 0.0120 & 0.0106 & 0.0107\\
sector/path mix & 0.0233 & 0.0216 & 0.0163 & 0.0185\\
barrier path & 0.0325 & 0.0254 & 0.0253 & 0.0311\\
\bottomrule
\end{tabular}
\end{table}

\section{Centered path-window barrier targets}

We next use \(n=8\) and compare a small set of non-diagonal target Hamiltonians whose geometry is expressed in the sector/path coordinates.  The target classes include original-window and \(v2\)-window barrier targets, sector/path mixture targets, and a sector-only well.  For each target, the baselines are fixed controls and the hybrid driver in \Eqref{eq:hybrid-driver} is optimized over the reproduced fine-scan grid in \(\alpha\) and \(\epsilon\) with path source original or \(v2\).  The purpose of this scan is not simply to find the largest number in one example.  It is to determine when the sector/path geometry helps and when it does not.

\begin{table}[htbp]
\centering
\caption{Regenerated \(n=8\), \(T=80\) non-diagonal target-class scan.  Baselines are fixed controls; the best-hybrid column is the best value on the reproduced target-class fine-scan grid.  This table uses that scan grid, whereas the \(0.9799\) value used in the convergence and ablation checks is the separate fixed-control \(w=8\) run.}
\label{tab:precise}
\small
\resizebox{\linewidth}{!}{%
\begin{tabular}{lcccccl}
\toprule
target & TF & sector & original path & \(v2\) path & best hybrid & best hybrid parameters\\
\midrule
original window barrier & 0.8902 & 0.9455 & 0.1739 & 0.7647 & 0.9704 & original, \((\alpha,\epsilon)=(0.30,0.10)\)\\
\(v2\) window barrier & 0.8553 & 0.9455 & 0.7353 & 0.1739 & 0.9688 & \(v2\), \((\alpha,\epsilon)=(0.25,0.10)\)\\
sector \(r=1\) sector well & 0.9985 & 0.9855 & 0.9107 & 0.9107 & 0.9904 & original, \((\alpha,\epsilon)=(0.20,0.10)\)\\
original mix sector/path & 0.9975 & 0.9881 & 0.7275 & 0.9015 & 0.9918 & \(v2\), \((\alpha,\epsilon)=(0.20,0.10)\)\\
\(v2\) mix sector/path & 0.9979 & 0.9881 & 0.9029 & 0.7275 & 0.9919 & original, \((\alpha,\epsilon)=(0.20,0.10)\)\\
\bottomrule
\end{tabular}%
}
\end{table}

\begin{figure}[htbp]
\centering
\includegraphics[width=.90\linewidth]{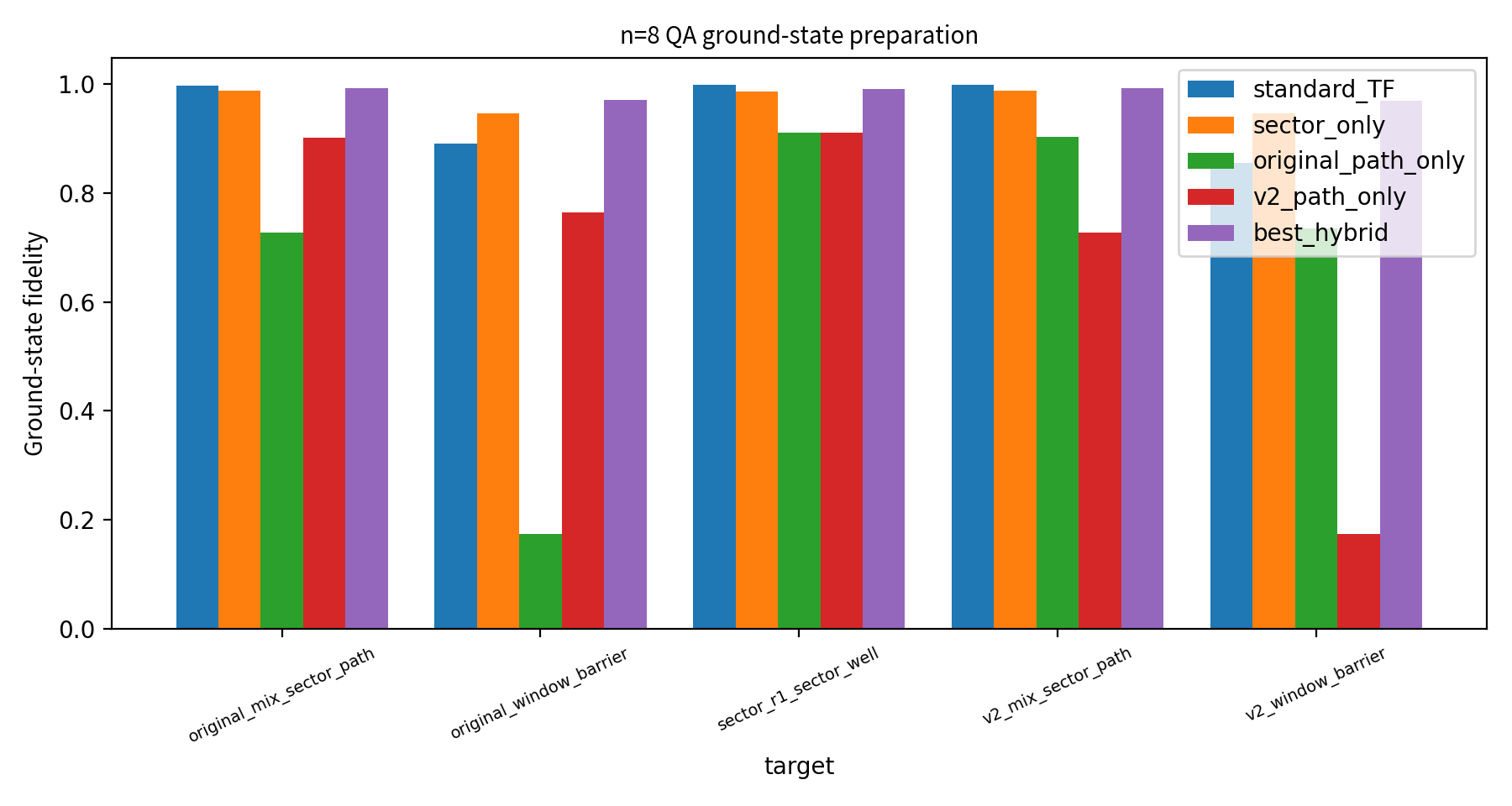}
\caption{Regenerated \(n=8\) baseline and best-hybrid comparison for the non-diagonal target classes in table~\ref{tab:precise}.  The sector-dominant hybrid improves the window-barrier targets, while standard TF remains very strong for sector/path mixture controls.}
\label{fig:precise}
\end{figure}

The regenerated scan supports a more conservative target-class statement.  The centered original-window barrier target gives transverse-field fidelity \(0.8902\), sector-only fidelity \(0.9455\), original path-only fidelity \(0.1739\), \(v2\) path-only fidelity \(0.7647\), and best available hybrid fidelity \(0.9704\) on the fine-scan grid.  A wider original-generator fixed-control check gives \(0.9799\) at \((w,\alpha,\epsilon)=(8,0.50,0.15)\).  These values support the qualitative conclusion that the centered barrier target is favored by sector-dominant hybrid graph drivers, while also showing that the precise numerical value depends on the scan grid and driver window.

Table~\ref{tab:precise} reports values obtained from the reproduced target-class fine-scan grid for each target class, whereas the convergence and ablation controls use fixed parameters unless explicitly stated otherwise.  In particular, the \(0.9704\) value in table~\ref{tab:precise} is the best centered original-window-barrier value on that target-class grid, while the \(0.9799\) value in the abstract, convergence table, and ablation controls is the separate fixed-control \(w=8\), \(\alpha=0.50\), \(\epsilon=0.15\) run from the wider original-generator check.  The single-component path-only controls also depend on the driver window: the original path-only value is \(0.1739\) for \(w=4\) and \(0.2490\) for \(w=8\).  These entries therefore should not be read as numerically inconsistent measurements of the same driver; they correspond to different fixed windows or scan grids.

\begin{figure}[htbp]
\centering
\includegraphics[width=.76\linewidth]{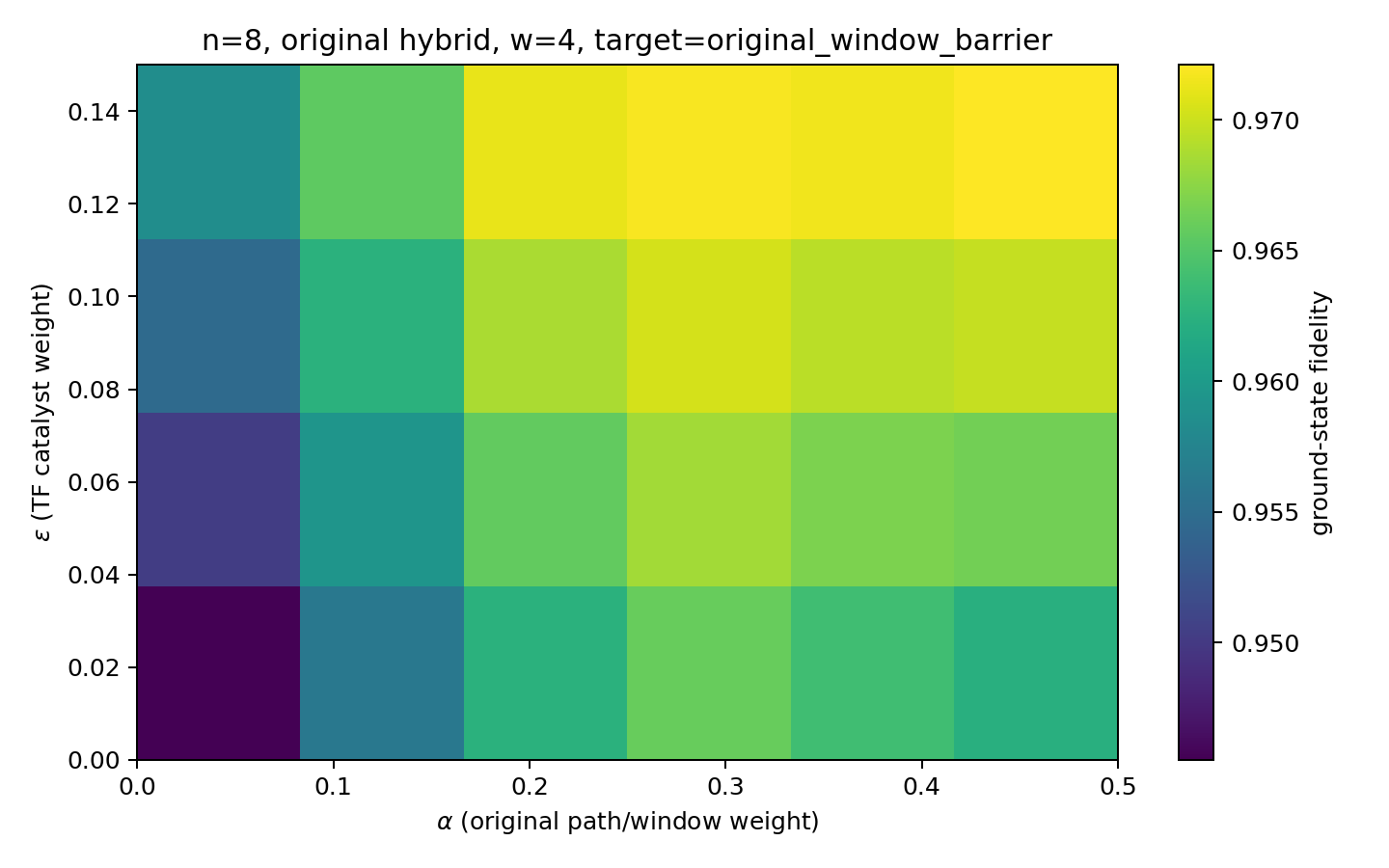}
\caption{Regenerated heatmap for the centered original-window barrier target with driver window \(w=4\).  The useful region is sector-dominant with a substantial but not standalone path-window component and a small transverse-field catalyst.}
\label{fig:heatmap}
\end{figure}

Figure~\ref{fig:heatmap} illustrates the role of the mixing parameters.  The best region is not \(\alpha=1\), which would be a path-only driver.  Instead, the high-fidelity region keeps a large sector component and uses the path-window term as an additional geometry-aware component.  This observation is the basis for the phrase ``sector-dominant hybrid driver'' used throughout the paper.

\topic{Status of the \(n=9\) strict attempt}
The exact strict-generator completion for \(n=9\) was attempted but did not complete within the attempted computation.  The corresponding timeout/status log is included in the reproduction archive, and no completed \(n=9\) strict-generator certificate is included or claimed.  Consequently, \(n=9\) is not used here as a completed strict-generator theorem or as evidence for an asymptotic scaling law; the quantitative QA conclusions are based on the regenerated finite \(n\le 8\) scans and the fixed-control checks summarized above.

\section{Controls, ablations, gaps, and finite-size behavior}

A natural objection is that the target was built from the strict ordering.  To test this, we compare matched path-order controls: for each source ordering, the target and the path-window component of the driver are constructed from that same source.  We also run cross-controls in which the target is fixed to the strict-ordering target but the path-window catalyst in the driver is changed.  These controls separate two questions: whether the strict ordering is a good matched geometry for its own target, and whether the improvement comes from a more general mechanism of adding a path-window catalyst to sector transport.

To strengthen the control study, we use explicit distributions over many sampled orderings rather than a single random example.  Two randomized ensembles are considered.  A random permutation is a fully random ordering of all \(2^8\) states.  A sector-preserving random ordering keeps the strict Hamming-weight skeleton \(\pi^{(8)}\) and the fixed initial strip, but fills the remaining slots by uniformly shuffled states of the required Hamming weight.  In both ensembles we sampled 64 orderings.  This sample size was chosen because each sample requires a full \(n=8\) finite-time evolution; at 64 samples the sector-preserving matched distribution already has a standard deviation of order \(10^{-3}\), so the qualitative conclusion is stable at the present level of precision.  The control parameters were kept at the deterministic values \(w=8\), \(\alpha=0.50\), and \(\epsilon=0.15\), so the distributions probe the path source rather than a separate parameter optimization.

\begin{table}[htbp]
\centering
\caption{Path-order controls at \(n=8\).  The matched columns use the same source ordering for target and driver.  The strict-target columns keep the centered strict target fixed and vary only the catalyst path in the driver.  Distribution rows report mean\(\pm\)standard deviation over 64 sampled orderings.}
\label{tab:controls}
\small
\resizebox{\linewidth}{!}{%
\begin{tabular}{lcc@{\qquad}cc}
\toprule
source / driver & matched fidelity & matched residual & strict-target fidelity & strict-target residual\\
\midrule
TF & -- & -- & 0.8902 & 0.0144\\
sector only & -- & -- & 0.9455 & 0.0140\\
strict/orig. & 0.9799 & 0.0085 & 0.9799 & 0.0085\\
Gray & 0.6718 & 0.0418 & 0.9617 & 0.0071\\
binary & 0.4754 & 0.0265 & 0.9482 & 0.0087\\
weight-block & 0.9594 & 0.0147 & 0.9534 & 0.0094\\
random perms. (64) & $0.3192\pm0.2770$ & $0.0470\pm0.0167$ & $0.9638\pm0.0049$ & $0.0100\pm0.0007$\\
sector-pres. random (64) & $0.9773\pm0.0010$ & $0.00864\pm0.00010$ & $0.9621\pm0.0065$ & $0.00837\pm0.00087$\\
\bottomrule
\end{tabular}%
}
\end{table}

\begin{figure}[htbp]
\centering
\includegraphics[width=\linewidth]{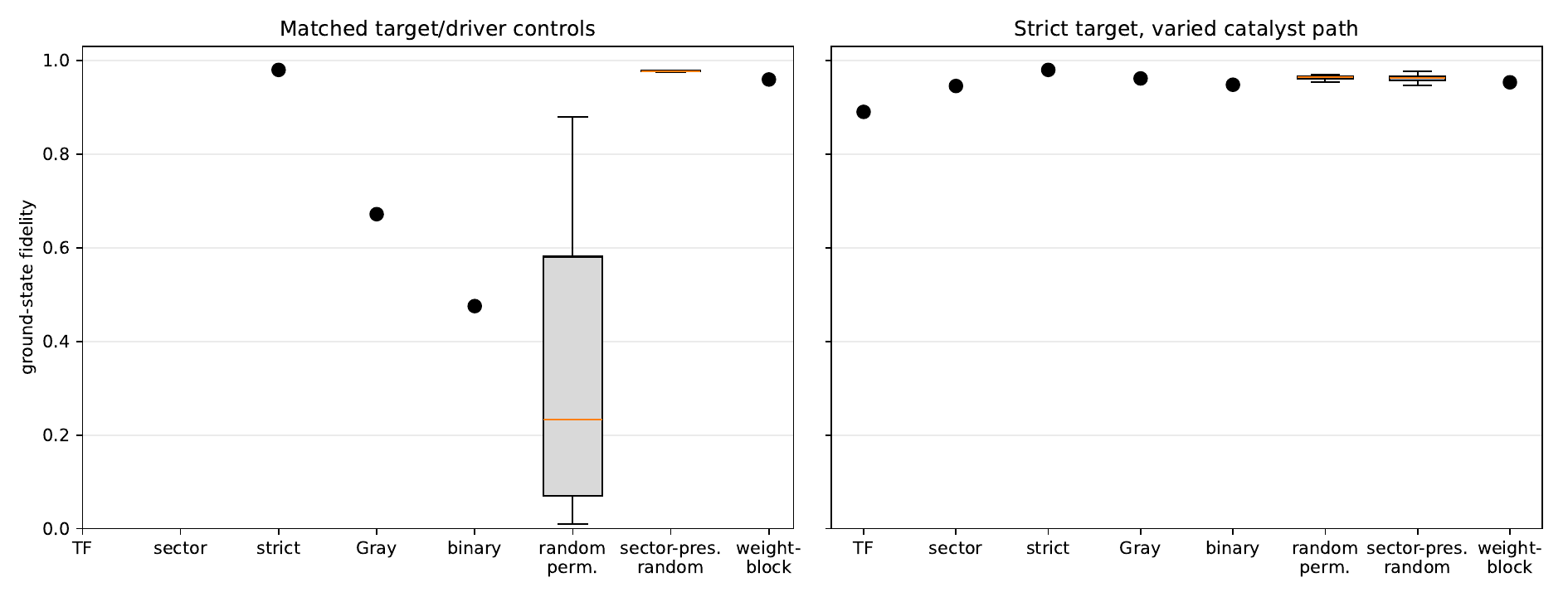}
\caption{Distributional path-order controls.  Boxplots show 64 sampled random permutations and 64 sampled sector-preserving random orderings; single points show TF, sector-only, and deterministic orderings.  Left: matched target/driver controls.  Right: strict target with varied catalyst path.}
\label{fig:controls}
\end{figure}

The distributional controls sharpen the interpretation.  Fully random permutations are poor in the matched setting, with mean fidelity only \(0.3192\pm0.2770\).  By contrast, sector-preserving random orderings are already excellent matched controls, with mean fidelity \(0.9773\pm0.0010\).  The data therefore do not suggest that the strict one-bit completion is uniquely responsible for the matched improvement.  Most of the effect already comes from the sector-preserving skeleton, while the strict completion provides at most a modest additional benefit in the tested instances.

In the strict-target cross-controls, many catalyst paths work well once the sector graph remains dominant.  The random-permutation and sector-preserving-random distributions have mean fidelities around \(0.96\), well above the sector-only baseline \(0.9455\), although still below the strict hybrid.  This confirms that the catalyst effect is genuinely broader than a single hand-picked path.  The safest interpretation is therefore twofold: there is a general sector-dominant plus path-window catalyst mechanism, and the strict sector-snake completion is one effective realization of that mechanism for the centered strict-window target used here.

To identify which driver components matter, we compare the full hybrid with several ablations on the centered strict-window target.  The path-window component alone is very weak.  The sector-only driver is already strong, and adding the path-window component and a small transverse-field component gives the best fidelity.

\begin{table}[htbp]
\centering
\caption{Driver ablation on the centered strict-window barrier target at \(n=8\), \(T=80\).}
\label{tab:ablation}
\begin{tabular}{lcc}
\toprule
driver & fidelity & energy residual\\
\midrule
sector + path + TF & 0.9799 & 0.0085\\
sector + path, no TF & 0.9697 & 0.0148\\
sector + TF, no path & 0.9585 & 0.0093\\
sector only & 0.9455 & 0.0140\\
TF only & 0.8902 & 0.0144\\
path + TF, no sector & 0.4614 & 0.1968\\
path only & 0.2490 & 0.3293\\
\bottomrule
\end{tabular}
\end{table}

\begin{figure}[htbp]
\centering
\includegraphics[width=\linewidth]{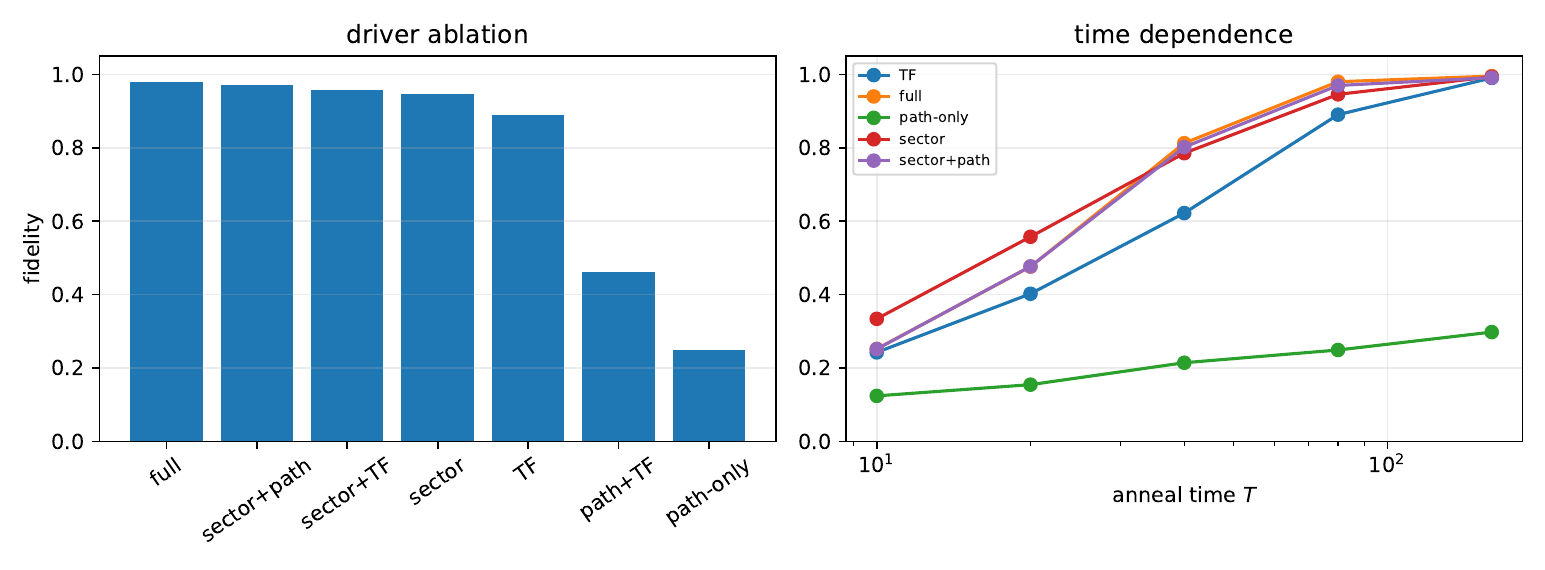}
\caption{Left: driver ablation.  The path-window graph is weak as a standalone driver, while the full sector-dominant hybrid is strongest.  Right: anneal-time dependence.  The full hybrid becomes clearly advantageous at intermediate and long anneal times, while all broad drivers approach high fidelity for \(T=160\).}
\label{fig:ablation-time}
\end{figure}

The ablation is one of the most important pieces of evidence in the paper.  It rules out a misleading interpretation in which the strict path itself is the good driver.  The strict path-window graph alone reaches fidelity only \(0.2490\), and even path plus transverse field without the sector graph remains poor.  The sector graph supplies the broad transport, and the path-window graph improves the match to the target only when embedded inside this broader transport graph.  Because \(\Gsec\) is much denser than \(\Gtf\), the comparison with TF-only should be read as a reference point rather than as a claim of equal-locality fairness.  The main mechanistic comparison is sector-only versus sector-plus-path and full hybrid.  On the centered \(n=8\) barrier target, adding the path-window catalyst raises the fidelity from \(0.9455\) to \(0.9697\), and adding the small transverse-field catalyst raises it further to \(0.9799\).

We also estimate spectral gaps on a grid in \(s\).  The path-only driver has an extremely small gap, consistent with its poor fidelity.  The full hybrid and sector-plus-path driver have their minimum grid gap at \(s=1\), equal to the target endpoint gap in this scan.  We therefore do not interpret the gap result as a proof that the hybrid opens a particular avoided crossing; it shows instead that the path-only small-gap pathology is avoided in the sector-dominant hybrid.

\begin{table}[htbp]
\centering
\caption{Grid-based minimum gaps on the centered original-window barrier target at \(n=8\) for selected regenerated controls.}
\label{tab:mingap}
\begin{tabular}{lcc}
\toprule
driver & \(s\) at minimum & minimum gap\\
\midrule
TF only & 0.9286 & 0.0690\\
sector only & 0.0000 & 0.0376\\
original window4 only & 0.0000 & 0.00118\\
original hybrid \(w=4,\alpha=0.30,\epsilon=0.10\) & 0.0000 & 0.0208\\
original hybrid \(w=8,\alpha=0.25,\epsilon=0.10\) & 1.0000 & 0.0691\\
\bottomrule
\end{tabular}
\end{table}

\begin{figure}[htbp]
\centering
\includegraphics[width=.49\linewidth]{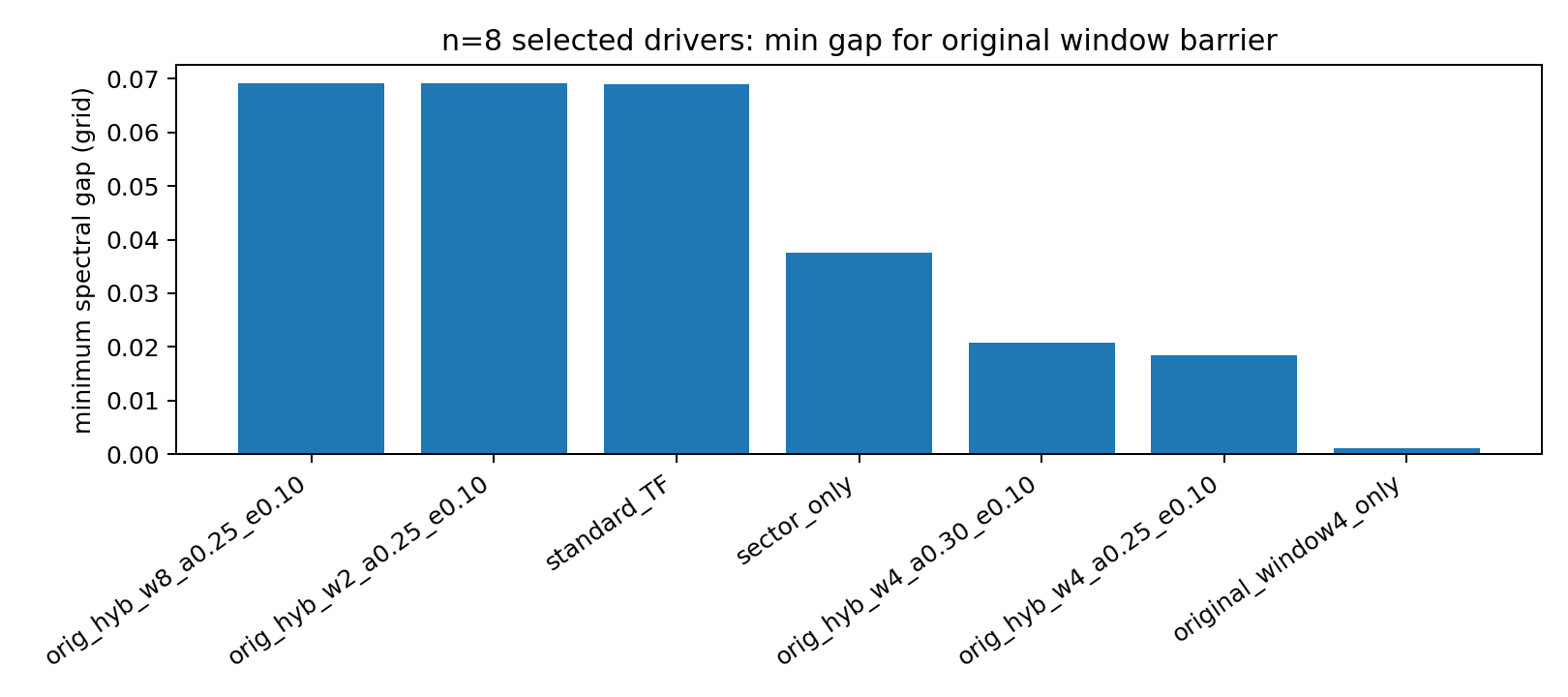}\hfill
\includegraphics[width=.49\linewidth]{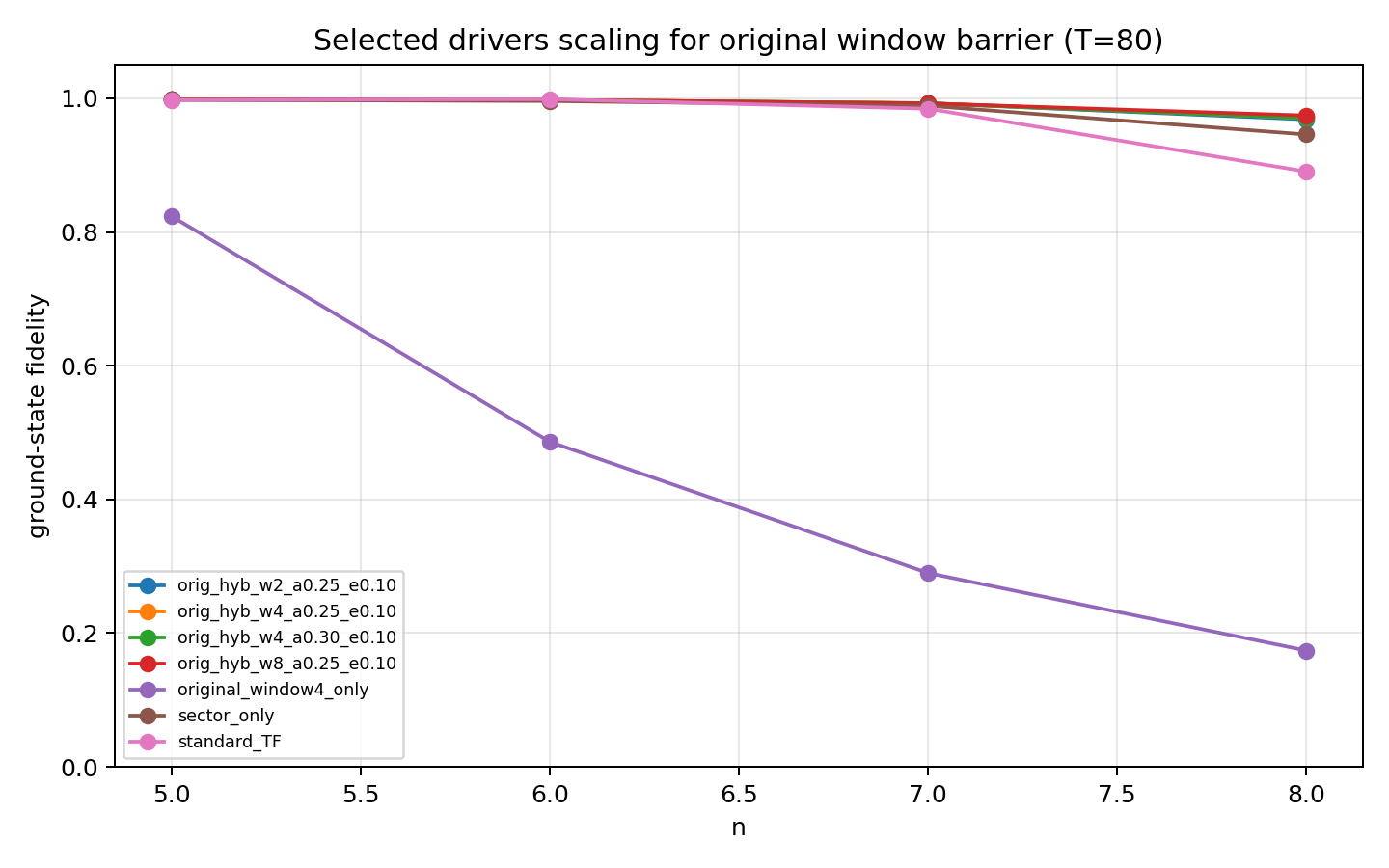}
\caption{Left: regenerated grid-based minimum gaps for selected drivers on the \(n=8\) original-window barrier target.  Right: finite-size summary for selected drivers over \(n=5,\ldots,8\) at \(T=80\).  The path-only driver deteriorates strongly, while sector-dominant hybrids remain high at the tested finite sizes.}
\label{fig:gap-scaling}
\end{figure}

The finite-size data for \(n=5,6,7,8\) should be read cautiously because \(T\) is fixed and the construction is finite-size.  Nevertheless, the trend is informative: the path-only driver degrades rapidly with \(n\), while the sector-dominant hybrid remains high at \(n=8\).  In small systems nearly all broad drivers can perform well; the difference becomes more visible at \(n=8\), where the centered barrier example is no longer trivial.

\section{Hamiltonian banding and an illustrative structured benchmark}

The preceding sections focus on adiabatic ground-state preparation.  The same ordering can also be tested as a matrix ordering for sparse Hamiltonians.  For a Hamiltonian with off-diagonal entries \(H_{xy}\), define the weighted mean band as
\begin{equation}
\operatorname{MeanBand}_{p}(H)=
\frac{\sum_{x<y}|H_{xy}|\,|p(x)-p(y)|}
     {\sum_{x<y}|H_{xy}|}.
\end{equation}
Smaller MeanBand means that the off-diagonal support is closer to the main diagonal after ordering the states by \(p\).  We also use OffBand measures in the data files, but MeanBand is the most compact comparison for the present paper.

This section contains two illustrative uses of the ordering that are deliberately kept separate because they use different metrics.  The first is Hamiltonian banding, for which the relevant metric is MeanBand.  The second is a structured staged sensor-placement relaxation, for which the relevant metric is finite-time ground-state-preparation fidelity.  We therefore do not combine these two uses into a single gain table in the main text.

Table~\ref{tab:banding-other-encodings} gives the detailed version of the Hamiltonian-banding experiment.  The compared orderings are the strict sector-snake ordering, the fast \(v2\) ordering, ordinary binary order, binary reflected Gray order, Hamming-weight block order, and 50 random permutations.  The random column reports a mean and standard deviation.  The table is deliberately included in full because it prevents overclaiming: the strict ordering is excellent for strict-path-local Hamiltonians but is not the best for every local Hamiltonian.

\begin{table}[htbp]
\centering
\caption{Hamiltonian MeanBand at \(n=8\) for several Hamiltonian families and several encodings.  The random column reports mean \(\pm\) standard deviation over 50 random permutations.  Smaller values are better.}
\label{tab:banding-other-encodings}
\small
\resizebox{\linewidth}{!}{%
\begin{tabular}{lrrrrrrl}
\toprule
Hamiltonian family & strict/orig. & \(v2\) & binary & Gray & weight-block & random & best deterministic\\
\midrule
sector dense \(r=1\) & 50.55 & 50.55 & 72.34 & 81.52 & 42.65 & \(85.72\pm0.83\) & weight-block\\
same-sector swap & 34.83 & 31.92 & 46.11 & 54.93 & 12.24 & \(85.76\pm1.28\) & weight-block\\
path-original \(w=4\) & \textbf{2.48} & 39.90 & 56.11 & 66.46 & 34.02 & \(85.72\pm1.85\) & strict/orig.\\
path-\(v2\) \(w=4\) & 43.78 & \textbf{2.48} & 44.92 & 51.42 & 28.79 & \(85.96\pm1.85\) & \(v2\)\\
mix sector+original path & \textbf{26.51} & 45.22 & 64.23 & 73.99 & 38.33 & \(85.72\pm1.04\) & strict/orig.\\
mix sector+\(v2\) path & 47.16 & \textbf{26.51} & 58.63 & 66.47 & 35.72 & \(85.84\pm1.08\) & \(v2\)\\
local hopping 1D & 33.51 & 22.22 & 18.14 & 36.29 & \textbf{4.67} & \(85.32\pm2.73\) & weight-block\\
local pair creation 1D & 92.44 & 89.28 & 54.43 & \textbf{36.29} & 98.51 & \(85.36\pm2.30\) & Gray\\
\bottomrule
\end{tabular}%
}
\end{table}

\begin{figure}[htbp]
\centering
\includegraphics[width=\linewidth]{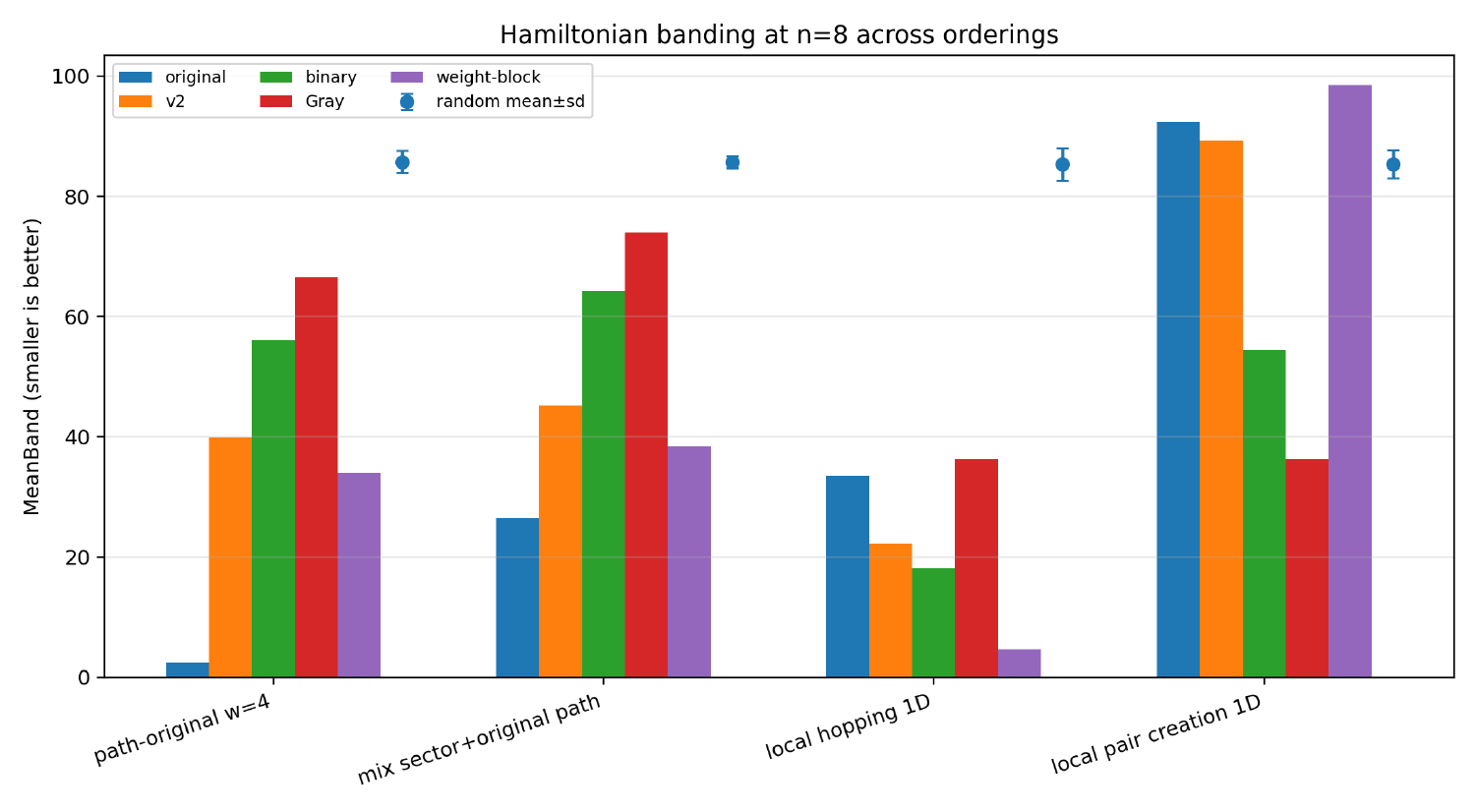}
\caption{Representative rows of table~\ref{tab:banding-other-encodings}.  The strict ordering is extremely effective for Hamiltonians whose off-diagonal support is local in the strict path coordinate.  It is not the best ordering for ordinary 1D hopping or pair creation, where a different locality is more natural.}
\label{fig:banding-other-encodings}
\end{figure}

The table makes the scope of the ordering advantage explicit.  When the Hamiltonian locality is defined by the strict path, the strict ordering reduces MeanBand from \(34.02\) for weight-block, \(39.90\) for \(v2\), \(56.11\) for binary order, and \(66.46\) for Gray order down to \(2.48\).  For the mixed sector-plus-original-path family, the strict ordering is again the best deterministic ordering in the table.  Conversely, for ordinary one-dimensional local hopping, weight-block is much better than the strict ordering, and for local pair creation Gray order is better.  Thus the ordering is not a generic bandwidth-reduction heuristic; it is useful when the Hamiltonian's off-diagonal locality is sector/path/window local.

As an illustrative structured benchmark, we next consider a small A-optimal sensor-placement problem.  There are \(n=8\) candidate sensor locations on a one-dimensional domain and 16 grid points at which a scalar field is to be estimated.  A subset \(S\subseteq\{1,\ldots,8\}\) represents the installed sensors.  The prior field covariance is a squared-exponential kernel
\begin{equation}
K(r,r')=\exp\left[-\frac{(r-r')^2}{2\ell^2}\right],
\qquad \ell=0.25,
\end{equation}
and the observation noise is \(\sigma=0.08\).  For a selected sensor set \(S\), the posterior covariance on the grid is
\begin{equation}
\Sigma_{\mathrm{post}}(S)
=
K_{gg}
-
K_{gS}
\left(K_{SS}+\sigma^2 I\right)^{-1}
K_{Sg}.
\end{equation}
We use the normalized A-optimality cost with a soft budget \(m=4\),
\begin{equation}
C_{\mathrm{sens}}(S)
=
\frac{\operatorname{tr}\Sigma_{\mathrm{post}}(S)}{\operatorname{tr}K_{gg}}
+
0.20\left(\frac{|S|-4}{8}\right)^2,
\end{equation}
rescaled to the interval \([0,1]\).

To test the sector/path encoding, we model staged deployment or reconfiguration by a path-window transition graph.  In this interpretation, two sensor subsets are close if they are close in the sector/path ordering and have nearby cardinality.  The target Hamiltonian is
\begin{equation}
H_T^{\mathrm{sens}}
=
0.6\,\widehat L(G_{E^{\mathrm{orig}},4})
+
\operatorname{diag}\bigl(C_{\mathrm{sens}}(S)\bigr).
\end{equation}
This is not an ordinary diagonal QUBO benchmark.  It is a graph-local relaxed sensor-placement Hamiltonian whose off-diagonal part represents allowed staged transitions between sensor configurations.  Therefore it is precisely the kind of problem for which a sector/path ordering can be relevant.

\begin{table}[htbp]
\centering
\caption{A-optimal sensor-placement benchmark at \(n=8\), \(T=80\).  The target is \(H_T^{\mathrm{sens}}=0.6\widehat L(G_{E^{\mathrm{orig}},4})+\operatorname{diag}(C_{\mathrm{sens}})\).}
\label{tab:sensor-placement}
\begin{tabular}{lccc}
\toprule
driver & \((\alpha,\epsilon)\) & fidelity & energy residual\\
\midrule
TF only & -- & 0.6655 & 0.0105\\
sector only & (0,0) & 0.7599 & 0.0107\\
path-window only & (1,0) & 0.2577 & 0.0542\\
sector+path, no TF & (0.50,0) & 0.7554 & 0.0067\\
hybrid & (0.30,0.10) & 0.8019 & 0.0068\\
hybrid & (0.50,0.20) & \textbf{0.8269} & \textbf{0.0051}\\
\bottomrule
\end{tabular}
\end{table}

\begin{figure}[htbp]
\centering
\includegraphics[width=.82\linewidth]{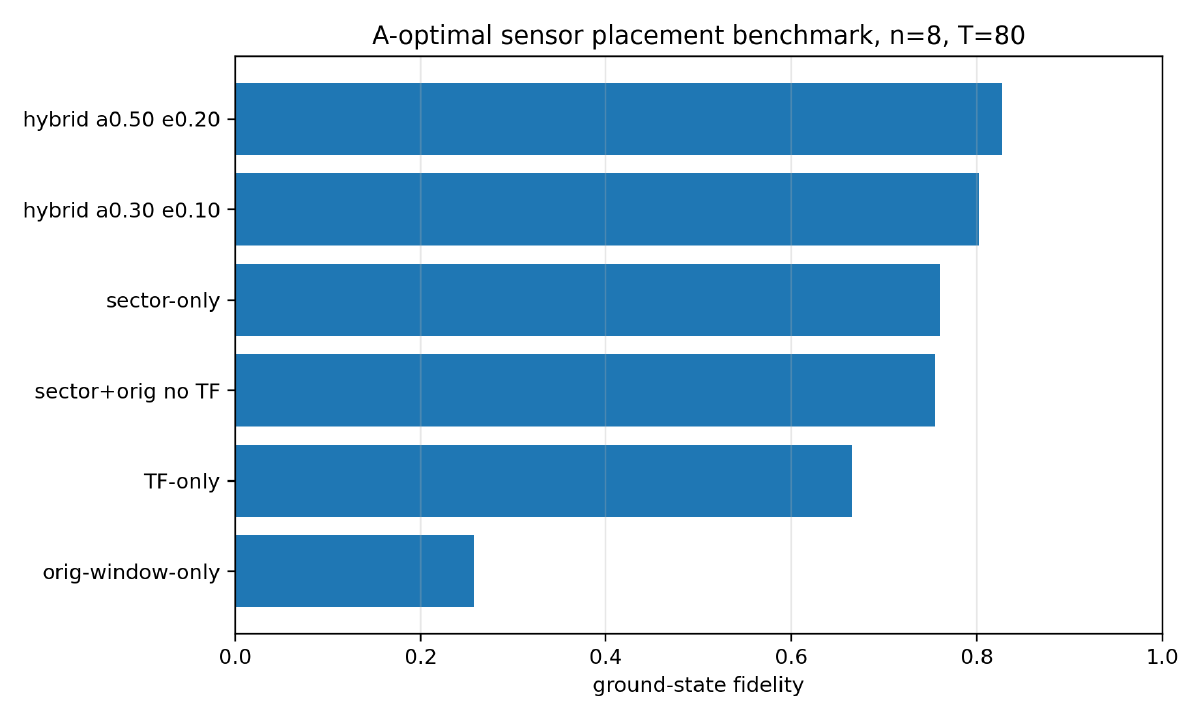}
\caption{Ground-state-preparation fidelity for the staged sensor-placement benchmark.  The sector-dominant hybrid driver improves over TF-only and sector-only, whereas the path-window driver alone remains weak.}
\label{fig:sensor-placement-qa}
\end{figure}

The staged sensor-placement calculation shows the same mechanism as the main barrier benchmark.  The best hybrid driver reaches fidelity \(0.8269\), compared with \(0.7599\) for sector-only and \(0.6655\) for TF-only.  The path-window driver alone is poor.  The result supports the proposed use case: the encoding is helpful when the practical problem supplies a transition or relaxation Hamiltonian that is local in the sector/path coordinate.  Similar interpretations are possible for staged feature selection, support search, or resource selection, provided that the transition graph has cardinality and path-window locality.

The matrix-ordering cases have a simple theoretical explanation related to classical matrix-bandwidth reduction ideas \cite{cuthill1969}.  If
\begin{equation}
    H_{xy}=0\quad\text{whenever}\quad |p_E(x)-p_E(y)|>w,
\end{equation}
then the matrix representation of \(H\) in the order \(p_E\) has bandwidth at most \(w\).  Thus, path-local Hamiltonians are exactly banded under their matched ordering.  This is a numerical-linear-algebra use of the ordering, distinct from adiabatic driver design.

\section{Discussion}

\topic{Intended purpose}
The role of the present paper is to formalize a reproducible finite-size ordering and to test what it is actually useful for.  The intended purpose is to make two coarse variables simultaneously available on the Boolean hypercube: the Hamming-weight sector \(\wt(x)\) and a path coordinate \(p_E(x)\).  When these two variables are meaningful for a problem, the ordering becomes more than a relabeling: it becomes a geometric object that can define driver graphs, barrier landscapes, and matrix orderings.

\topic{What is established and what is not}
The strict sector-snake ordering is a fixed-prefix representative of a monotone Gray code that respects Hamming-weight sectors.  Its most reliable use is not as a direct relabeling for standard transverse-field QA, but as a way to define a path coordinate and path-window graph.  When the target Hamiltonian is non-diagonal and local in this graph, a hybrid driver can exploit the additional geometry.  The sector graph is the main transport component; the path-window graph is a catalyst that is useful when aligned with the target; and the small transverse-field component gives additional hypercube-local flexibility.  Because the sector graph is much denser than the transverse-field graph, the comparison with TF-only is best read as a reference baseline.  The mechanistic comparison that matters most is sector-only versus sector-plus-path and full hybrid.

Because the non-diagonal target Hamiltonians are indexed by the same sector/path coordinate used in the driver construction, the relevant question is not coordinate alignment alone, but robustness under matched controls, strict-target cross-controls, and sector-preserving randomizations.  The numerical data indicate that the dominant contribution comes from the sector-preserving skeleton, while the strict one-bit completion acts as a secondary refinement.  We therefore do not claim that the strict ordering itself is a universal improvement mechanism; rather, it provides a controlled representative for testing sector/path-local graph drivers.

The simulations do not support the claim that the strict ordering gives a universal advantage for ordinary diagonal QUBO-style annealing.  The data also do not support using the path-window graph alone as a driver.  A thin path or narrow window behaves too much like a one-dimensional chain and can have a very small spectral gap.  Finally, the hybrid driver is a custom graph-local driver, not a direct implementation on present-day transverse-field Ising annealing hardware.

\topic{Structured quantum state preparation}
One natural motivation for the present construction is adiabatic quantum state preparation.  In that setting, the final ground state is itself the desired many-body or entangled state rather than merely the minimizer of a diagonal classical cost, and the choice of interpolation path can be crucial \cite{farhi2000,babbush2014asp}.  Our centered barrier simulations fit exactly this interpretation: \(H_T\) is a structured model Hamiltonian, \(H_D\) is a transport Hamiltonian, and the question is whether a sector-dominant hybrid path can improve finite-time preparation.  The answer supported by the data is yes for centered path-window barrier targets.  What is \emph{not} established here is an asymptotic speedup theorem, a proof that exponentially slow cases become polynomial, or a general robustness claim against experimental noise.  The correct statement is a finite-size design principle for structured non-diagonal ground-state preparation.

\topic{Coarse-grained folding and conformational-search interpretation}
A second plausible interpretation is coarse-grained folding or conformational search.  One may let \(x_i=1\) mean that contact \(i\), motif \(i\), or local conformer \(i\) is formed; then \(\wt(x)\) is a contact-count or compactness coordinate, and \(p_E(x)\) is a discrete reaction coordinate.  A barrier target \(H_T=\widehat L(G_{E,w})+V\) can then be read as a toy funnel-like landscape with intermediate kinetic traps \cite{wolynes1997,perdomo2012}.  Within this interpretation the sector graph allows lateral rearrangements among states of similar compactness, while the path-window component biases the motion toward an intended route.  This makes the model a reasonable structured test bed for folding-like search problems.  At the same time, the present paper does not simulate realistic protein force fields and does not justify strong application claims such as direct atomistic folding prediction or immediate drug-discovery speedups.

\topic{What the encoding can concretely be used for}
The uses actually supported by the current numerics fall into two groups.  The first is graph-local adiabatic state preparation and relaxed subset optimization, as illustrated by the centered barrier targets and the staged sensor-placement benchmark.  The second is matrix ordering for sparse Hamiltonians and diffusion generators, as illustrated by the MeanBand and band-truncation calculations.  Prospective applications include staged feature selection, support search, defect or dopant configuration models, and subset-space Markov dynamics whenever the transition graph has sector/path locality.  The key point is that the encoding is useful when the problem supplies a non-diagonal transport graph or matrix locality compatible with \(\wt(x)\) and \(p_E(x)\).

\topic{Limitations and positioning}
There are three important limitations.  First, the monotone Gray-code class itself is not new, and all-\(n\) existence of some monotone Gray code is known.  What is used here is a finite representative-selection procedure checked through the regenerated cases; the strict \(n=8\) path is validated, while the strict \(n=9\) attempt did not complete within the attempted computation.  The all-\(n\) success of our representative-selection rule remains unproved.  Second, the ordering controls use finite sampled distributions rather than exhaustive characterizations of all compatible paths.  Third, the gap estimates are grid-based and, in the hybrid cases, the minimum can occur at the endpoint.  We therefore interpret the present simulations as finite-time ground-state-preparation evidence rather than as an asymptotic gap theorem.

These limitations also clarify how the work should be positioned.  The paper is best read as a structured numerical and conceptual study of graph-local drivers.  It identifies a class of Hamiltonians for which the sector/path ordering is a useful design tool, and it identifies several classes for which it is not.  This separation is important for avoiding an overly broad claim and for making the result useful to readers interested in problem-dependent adiabatic models.

\section{Concluding remarks}

In this article, we have used a fixed-prefix sector-snake representative of the monotone Gray-code class as a sector/path coordinate on the Boolean hypercube, and examined what it is actually useful for.  The monotone Gray-code existence problem is not new: the relevant all-\(n\) existence theory is due to Savage and Winkler.  The finite construction reported here is instead a concrete representative-selection procedure: the \(n\le 8\) paths arise from deterministic completion and the strict \(n=8\) path is validated by the compiled generator log, while the strict \(n=9\) attempt did not complete within the attempted computation.  The coordinate was introduced to expose Hamming-weight sector structure, provide a discrete path coordinate, and organize path-local matrices on the Boolean hypercube.  Direct use as a universal encoding for standard transverse-field diagonal QA is not supported.  However, when the coordinate is used to define path-window graph locality, it becomes useful in combination with a broad Hamming-weight sector graph.  For centered path-window barrier targets, the regenerated \(n=8\) scans and fixed-control checks show that a sector-dominant hybrid graph-local driver can outperform both the standard transverse-field driver and the sector-only driver in the tested finite instances.  Ablation and controls indicate that the path-window component acts as a catalyst rather than a standalone driver.  The resulting proposal is therefore conditional and structured: sector/path coordinates are useful for designing graph-local drivers, coarse reaction-coordinate models, and banded Hamiltonian representations when the problem Hamiltonian itself has sector/path/window locality.

\section*{Data availability statement}
The source code, completed finite certificates, CSV files, validation logs, timeout logs, and reproduction scripts used in this work are supplied as the accompanying ancillary/supplementary reproduction package for this version.  The package includes the completed \(n=8\) strict-generator certificate and validation log, the \(n=9\) timeout/status log, regenerated CSV summaries, figure-generation scripts, and repository-ready rerun helpers.  No completed \(n=9\) strict-generator certificate is included or claimed.  The same package can be used as journal supplementary material; a public GitHub or Zenodo DOI can be cited in a later version if finalized.

\section*{Appendix. Finite-size validation, practical examples, and proof of the band-locality statement}
\addcontentsline{toc}{section}{Appendix. Finite-size validation, practical examples, and proof of the band-locality statement}

\medskip
\noindent\textbf{Finite-size validation status.}
The reproduction package includes the completed \(n=8\) strict-generator path and validation log, together with the \(n=9\) strict-generator timeout/status log.  The \(n=8\) log verifies a path of length \(256\).  The \(n=9\) log records that the attempted strict completion did not finish within the attempted computation; in the regenerated log it stopped after about \(50.001\) seconds with \(972{,}029{,}952\) searched nodes and partial path index \(479\).  These files make the finite-size status of the selected representative explicit without turning the attempted \(n=9\) computation into a completed certificate.

The abstract Hamiltonian classes discussed in the main text can be mapped to concrete structured problems.  The staged sensor-placement benchmark in the main text is a numerical demonstration of this idea.  Other examples are prospective mappings.  We keep these examples in the appendix because their natural performance metrics differ---some are adiabatic fidelities, while others are matrix-ordering quantities such as MeanBand---and we do not want to combine unlike metrics in a single headline table.  A binary contact-folding model uses \(x_i=1\) to indicate that contact \(i\) is formed; \(\wt(x)\) is the number of formed contacts and \(p_E(x)\) is a reaction coordinate.  A defect or dopant configuration model uses \(x_i=1\) to indicate an occupied site; \([j:k]\)-local Hamiltonians describe transitions between nearby defect-count sectors and nearby block indices.  A sensor-placement or feature-selection problem uses \(x_i=1\) for selected sensors or features and naturally contains cardinality penalties such as \((\wt(x)-m)^2\).  A Markov or diffusion model on subsets uses a graph Laplacian to describe transitions between nearby configurations and studies low eigenmodes or metastability.  These interpretations should be read as coarse-grained structured mappings, not as completed application claims.  Their relevance depends on whether the physical or algorithmic transition graph really has sector/path locality; otherwise, the sector/path ordering is not expected to help.

\medskip
\noindent\textbf{Proposition A.1.} Let \(H\) be a matrix indexed by \(Q_n\). If
\begin{equation}
    H_{xy}=0\quad\text{whenever}\quad |p_E(x)-p_E(y)|>w,
\end{equation}
then the matrix representation of \(H\) in the ordering \(p_E\) has bandwidth at most \(w\).

\medskip
\noindent\textit{Proof.} The row and column indices of the permuted matrix are \(p_E(x)\) and \(p_E(y)\). A non-zero off-diagonal entry can occur only if \(|p_E(x)-p_E(y)|\le w\), by assumption. Hence every non-zero entry lies within \(w\) diagonals of the main diagonal, which is exactly the definition of bandwidth at most \(w\). \(\square\)

The numerical MeanBand and OffBand experiments are therefore best understood as tests of whether the chosen ordering matches the off-diagonal locality of a given Hamiltonian.

\medskip
\noindent\textbf{Explicit sector-snake encodings for \(n=5,6,7,8\).}
The following tables list the canonical strict encodings used in the numerical experiments for \(n=5,6,7,8\).  A subset string such as \(134\) means \(\{1,3,4\}\), and \(\emptyset\) denotes the empty set.  The bitstring \(E_n(t)\) is printed in the convention used throughout the code: the leftmost bit corresponds to element \(n\), and the rightmost bit corresponds to element \(1\).  These tables are included to make the proposed finite-size encoding fully reproducible.

\begingroup
\scriptsize
\setlength{\tabcolsep}{3pt}
\renewcommand{\arraystretch}{0.92}
\begin{longtable}{@{}rll rll rll @{}}
\caption{Explicit strict sector-snake encoding for $n=5$.  Each entry gives $t$, subset $S_t$, and bitstring $E_5(t)$.}\label{tab:encoding-n5}\\
\toprule
$t$ & $S_t$ & $E_5(t)$ & $t$ & $S_t$ & $E_5(t)$ & $t$ & $S_t$ & $E_5(t)$\\
\midrule
\endfirsthead
\caption[]{Explicit strict sector-snake encoding for $n=5$ (continued).}\\
\toprule
$t$ & $S_t$ & $E_5(t)$ & $t$ & $S_t$ & $E_5(t)$ & $t$ & $S_t$ & $E_5(t)$\\
\midrule
\endhead
0 & $\emptyset$ & \texttt{00000} & 11 & $235$ & \texttt{10110} & 22 & $1345$ & \texttt{11101}\\
1 & $1$ & \texttt{00001} & 12 & $25$ & \texttt{10010} & 23 & $345$ & \texttt{11100}\\
2 & $12$ & \texttt{00011} & 13 & $245$ & \texttt{11010} & 24 & $2345$ & \texttt{11110}\\
3 & $2$ & \texttt{00010} & 14 & $24$ & \texttt{01010} & 25 & $234$ & \texttt{01110}\\
4 & $23$ & \texttt{00110} & 15 & $124$ & \texttt{01011} & 26 & $1234$ & \texttt{01111}\\
5 & $3$ & \texttt{00100} & 16 & $14$ & \texttt{01001} & 27 & $123$ & \texttt{00111}\\
6 & $34$ & \texttt{01100} & 17 & $145$ & \texttt{11001} & 28 & $1235$ & \texttt{10111}\\
7 & $4$ & \texttt{01000} & 18 & $15$ & \texttt{10001} & 29 & $125$ & \texttt{10011}\\
8 & $45$ & \texttt{11000} & 19 & $135$ & \texttt{10101} & 30 & $1245$ & \texttt{11011}\\
9 & $5$ & \texttt{10000} & 20 & $13$ & \texttt{00101} & 31 & $12345$ & \texttt{11111}\\
10 & $35$ & \texttt{10100} & 21 & $134$ & \texttt{01101} &  &  & \\
\bottomrule
\end{longtable}
\endgroup

\begingroup
\scriptsize
\setlength{\tabcolsep}{3pt}
\renewcommand{\arraystretch}{0.92}
\begin{longtable}{@{}rll rll rll @{}}
\caption{Explicit strict sector-snake encoding for $n=6$.  Each entry gives $t$, subset $S_t$, and bitstring $E_6(t)$.}\label{tab:encoding-n6}\\
\toprule
$t$ & $S_t$ & $E_6(t)$ & $t$ & $S_t$ & $E_6(t)$ & $t$ & $S_t$ & $E_6(t)$\\
\midrule
\endfirsthead
\caption[]{Explicit strict sector-snake encoding for $n=6$ (continued).}\\
\toprule
$t$ & $S_t$ & $E_6(t)$ & $t$ & $S_t$ & $E_6(t)$ & $t$ & $S_t$ & $E_6(t)$\\
\midrule
\endhead
0 & $\emptyset$ & \texttt{000000} & 22 & $24$ & \texttt{001010} & 44 & $2346$ & \texttt{101110}\\
1 & $1$ & \texttt{000001} & 23 & $124$ & \texttt{001011} & 45 & $236$ & \texttt{100110}\\
2 & $12$ & \texttt{000011} & 24 & $14$ & \texttt{001001} & 46 & $1236$ & \texttt{100111}\\
3 & $2$ & \texttt{000010} & 25 & $146$ & \texttt{101001} & 47 & $126$ & \texttt{100011}\\
4 & $23$ & \texttt{000110} & 26 & $16$ & \texttt{100001} & 48 & $1256$ & \texttt{110011}\\
5 & $3$ & \texttt{000100} & 27 & $156$ & \texttt{110001} & 49 & $125$ & \texttt{010011}\\
6 & $34$ & \texttt{001100} & 28 & $15$ & \texttt{010001} & 50 & $1235$ & \texttt{010111}\\
7 & $4$ & \texttt{001000} & 29 & $135$ & \texttt{010101} & 51 & $123$ & \texttt{000111}\\
8 & $45$ & \texttt{011000} & 30 & $13$ & \texttt{000101} & 52 & $1234$ & \texttt{001111}\\
9 & $5$ & \texttt{010000} & 31 & $136$ & \texttt{100101} & 53 & $12346$ & \texttt{101111}\\
10 & $56$ & \texttt{110000} & 32 & $1346$ & \texttt{101101} & 54 & $1246$ & \texttt{101011}\\
11 & $6$ & \texttt{100000} & 33 & $134$ & \texttt{001101} & 55 & $12456$ & \texttt{111011}\\
12 & $46$ & \texttt{101000} & 34 & $1345$ & \texttt{011101} & 56 & $2456$ & \texttt{111010}\\
13 & $346$ & \texttt{101100} & 35 & $345$ & \texttt{011100} & 57 & $23456$ & \texttt{111110}\\
14 & $36$ & \texttt{100100} & 36 & $3456$ & \texttt{111100} & 58 & $2356$ & \texttt{110110}\\
15 & $356$ & \texttt{110100} & 37 & $456$ & \texttt{111000} & 59 & $12356$ & \texttt{110111}\\
16 & $35$ & \texttt{010100} & 38 & $1456$ & \texttt{111001} & 60 & $1356$ & \texttt{110101}\\
17 & $235$ & \texttt{010110} & 39 & $145$ & \texttt{011001} & 61 & $13456$ & \texttt{111101}\\
18 & $25$ & \texttt{010010} & 40 & $1245$ & \texttt{011011} & 62 & $123456$ & \texttt{111111}\\
19 & $256$ & \texttt{110010} & 41 & $245$ & \texttt{011010} & 63 & $12345$ & \texttt{011111}\\
20 & $26$ & \texttt{100010} & 42 & $2345$ & \texttt{011110} &  &  & \\
21 & $246$ & \texttt{101010} & 43 & $234$ & \texttt{001110} &  &  & \\
\bottomrule
\end{longtable}
\endgroup

\begingroup
\scriptsize
\setlength{\tabcolsep}{3pt}
\renewcommand{\arraystretch}{0.92}
\begin{longtable}{@{}rll rll rll @{}}
\caption{Explicit strict sector-snake encoding for $n=7$.  Each entry gives $t$, subset $S_t$, and bitstring $E_7(t)$.}\label{tab:encoding-n7}\\
\toprule
$t$ & $S_t$ & $E_7(t)$ & $t$ & $S_t$ & $E_7(t)$ & $t$ & $S_t$ & $E_7(t)$\\
\midrule
\endfirsthead
\caption[]{Explicit strict sector-snake encoding for $n=7$ (continued).}\\
\toprule
$t$ & $S_t$ & $E_7(t)$ & $t$ & $S_t$ & $E_7(t)$ & $t$ & $S_t$ & $E_7(t)$\\
\midrule
\endhead
0 & $\emptyset$ & \texttt{0000000} & 43 & $137$ & \texttt{1000101} & 86 & $2467$ & \texttt{1101010}\\
1 & $1$ & \texttt{0000001} & 44 & $1367$ & \texttt{1100101} & 87 & $23467$ & \texttt{1101110}\\
2 & $12$ & \texttt{0000011} & 45 & $136$ & \texttt{0100101} & 88 & $3467$ & \texttt{1101100}\\
3 & $2$ & \texttt{0000010} & 46 & $1356$ & \texttt{0110101} & 89 & $13467$ & \texttt{1101101}\\
4 & $23$ & \texttt{0000110} & 47 & $356$ & \texttt{0110100} & 90 & $1467$ & \texttt{1101001}\\
5 & $3$ & \texttt{0000100} & 48 & $3567$ & \texttt{1110100} & 91 & $14567$ & \texttt{1111001}\\
6 & $34$ & \texttt{0001100} & 49 & $567$ & \texttt{1110000} & 92 & $1567$ & \texttt{1110001}\\
7 & $4$ & \texttt{0001000} & 50 & $4567$ & \texttt{1111000} & 93 & $13567$ & \texttt{1110101}\\
8 & $45$ & \texttt{0011000} & 51 & $456$ & \texttt{0111000} & 94 & $1357$ & \texttt{1010101}\\
9 & $5$ & \texttt{0010000} & 52 & $3456$ & \texttt{0111100} & 95 & $13457$ & \texttt{1011101}\\
10 & $56$ & \texttt{0110000} & 53 & $345$ & \texttt{0011100} & 96 & $1345$ & \texttt{0011101}\\
11 & $6$ & \texttt{0100000} & 54 & $3457$ & \texttt{1011100} & 97 & $12345$ & \texttt{0011111}\\
12 & $67$ & \texttt{1100000} & 55 & $347$ & \texttt{1001100} & 98 & $1235$ & \texttt{0010111}\\
13 & $7$ & \texttt{1000000} & 56 & $1347$ & \texttt{1001101} & 99 & $12356$ & \texttt{0110111}\\
14 & $57$ & \texttt{1010000} & 57 & $134$ & \texttt{0001101} & 100 & $1236$ & \texttt{0100111}\\
15 & $457$ & \texttt{1011000} & 58 & $1346$ & \texttt{0101101} & 101 & $12367$ & \texttt{1100111}\\
16 & $47$ & \texttt{1001000} & 59 & $146$ & \texttt{0101001} & 102 & $2367$ & \texttt{1100110}\\
17 & $467$ & \texttt{1101000} & 60 & $1456$ & \texttt{0111001} & 103 & $23567$ & \texttt{1110110}\\
18 & $46$ & \texttt{0101000} & 61 & $145$ & \texttt{0011001} & 104 & $2357$ & \texttt{1010110}\\
19 & $346$ & \texttt{0101100} & 62 & $1457$ & \texttt{1011001} & 105 & $23457$ & \texttt{1011110}\\
20 & $36$ & \texttt{0100100} & 63 & $157$ & \texttt{1010001} & 106 & $2345$ & \texttt{0011110}\\
21 & $367$ & \texttt{1100100} & 64 & $1257$ & \texttt{1010011} & 107 & $23456$ & \texttt{0111110}\\
22 & $37$ & \texttt{1000100} & 65 & $127$ & \texttt{1000011} & 108 & $2456$ & \texttt{0111010}\\
23 & $357$ & \texttt{1010100} & 66 & $1267$ & \texttt{1100011} & 109 & $12456$ & \texttt{0111011}\\
24 & $35$ & \texttt{0010100} & 67 & $126$ & \texttt{0100011} & 110 & $1246$ & \texttt{0101011}\\
25 & $235$ & \texttt{0010110} & 68 & $1256$ & \texttt{0110011} & 111 & $12467$ & \texttt{1101011}\\
26 & $25$ & \texttt{0010010} & 69 & $125$ & \texttt{0010011} & 112 & $1247$ & \texttt{1001011}\\
27 & $257$ & \texttt{1010010} & 70 & $1245$ & \texttt{0011011} & 113 & $12457$ & \texttt{1011011}\\
28 & $27$ & \texttt{1000010} & 71 & $245$ & \texttt{0011010} & 114 & $124567$ & \texttt{1111011}\\
29 & $267$ & \texttt{1100010} & 72 & $2457$ & \texttt{1011010} & 115 & $12567$ & \texttt{1110011}\\
30 & $26$ & \texttt{0100010} & 73 & $247$ & \texttt{1001010} & 116 & $123567$ & \texttt{1110111}\\
31 & $246$ & \texttt{0101010} & 74 & $2347$ & \texttt{1001110} & 117 & $12357$ & \texttt{1010111}\\
32 & $24$ & \texttt{0001010} & 75 & $237$ & \texttt{1000110} & 118 & $123457$ & \texttt{1011111}\\
33 & $124$ & \texttt{0001011} & 76 & $1237$ & \texttt{1000111} & 119 & $12347$ & \texttt{1001111}\\
34 & $14$ & \texttt{0001001} & 77 & $123$ & \texttt{0000111} & 120 & $123467$ & \texttt{1101111}\\
35 & $147$ & \texttt{1001001} & 78 & $1234$ & \texttt{0001111} & 121 & $12346$ & \texttt{0101111}\\
36 & $17$ & \texttt{1000001} & 79 & $234$ & \texttt{0001110} & 122 & $123456$ & \texttt{0111111}\\
37 & $167$ & \texttt{1100001} & 80 & $2346$ & \texttt{0101110} & 123 & $13456$ & \texttt{0111101}\\
38 & $16$ & \texttt{0100001} & 81 & $236$ & \texttt{0100110} & 124 & $134567$ & \texttt{1111101}\\
39 & $156$ & \texttt{0110001} & 82 & $2356$ & \texttt{0110110} & 125 & $34567$ & \texttt{1111100}\\
40 & $15$ & \texttt{0010001} & 83 & $256$ & \texttt{0110010} & 126 & $234567$ & \texttt{1111110}\\
41 & $135$ & \texttt{0010101} & 84 & $2567$ & \texttt{1110010} & 127 & $1234567$ & \texttt{1111111}\\
42 & $13$ & \texttt{0000101} & 85 & $24567$ & \texttt{1111010} &  &  & \\
\bottomrule
\end{longtable}
\endgroup

\begingroup
\scriptsize
\setlength{\tabcolsep}{3pt}
\renewcommand{\arraystretch}{0.92}
\begin{longtable}{@{}rll rll rll @{}}
\caption{Explicit strict sector-snake encoding for $n=8$.  Each entry gives $t$, subset $S_t$, and bitstring $E_8(t)$.}\label{tab:encoding-n8}\\
\toprule
$t$ & $S_t$ & $E_8(t)$ & $t$ & $S_t$ & $E_8(t)$ & $t$ & $S_t$ & $E_8(t)$\\
\midrule
\endfirsthead
\caption[]{Explicit strict sector-snake encoding for $n=8$ (continued).}\\
\toprule
$t$ & $S_t$ & $E_8(t)$ & $t$ & $S_t$ & $E_8(t)$ & $t$ & $S_t$ & $E_8(t)$\\
\midrule
\endhead
0 & $\emptyset$ & \texttt{00000000} & 86 & $4568$ & \texttt{10111000} & 172 & $1368$ & \texttt{10100101}\\
1 & $1$ & \texttt{00000001} & 87 & $456$ & \texttt{00111000} & 173 & $13468$ & \texttt{10101101}\\
2 & $12$ & \texttt{00000011} & 88 & $3456$ & \texttt{00111100} & 174 & $3468$ & \texttt{10101100}\\
3 & $2$ & \texttt{00000010} & 89 & $356$ & \texttt{00110100} & 175 & $23468$ & \texttt{10101110}\\
4 & $23$ & \texttt{00000110} & 90 & $3568$ & \texttt{10110100} & 176 & $2348$ & \texttt{10001110}\\
5 & $3$ & \texttt{00000100} & 91 & $358$ & \texttt{10010100} & 177 & $23478$ & \texttt{11001110}\\
6 & $34$ & \texttt{00001100} & 92 & $2358$ & \texttt{10010110} & 178 & $2347$ & \texttt{01001110}\\
7 & $4$ & \texttt{00001000} & 93 & $238$ & \texttt{10000110} & 179 & $23457$ & \texttt{01011110}\\
8 & $45$ & \texttt{00011000} & 94 & $2378$ & \texttt{11000110} & 180 & $2357$ & \texttt{01010110}\\
9 & $5$ & \texttt{00010000} & 95 & $237$ & \texttt{01000110} & 181 & $12357$ & \texttt{01010111}\\
10 & $56$ & \texttt{00110000} & 96 & $2367$ & \texttt{01100110} & 182 & $1237$ & \texttt{01000111}\\
11 & $6$ & \texttt{00100000} & 97 & $367$ & \texttt{01100100} & 183 & $12347$ & \texttt{01001111}\\
12 & $67$ & \texttt{01100000} & 98 & $3467$ & \texttt{01101100} & 184 & $1247$ & \texttt{01001011}\\
13 & $7$ & \texttt{01000000} & 99 & $347$ & \texttt{01001100} & 185 & $12467$ & \texttt{01101011}\\
14 & $78$ & \texttt{11000000} & 100 & $3478$ & \texttt{11001100} & 186 & $1246$ & \texttt{00101011}\\
15 & $8$ & \texttt{10000000} & 101 & $348$ & \texttt{10001100} & 187 & $12468$ & \texttt{10101011}\\
16 & $68$ & \texttt{10100000} & 102 & $3458$ & \texttt{10011100} & 188 & $1468$ & \texttt{10101001}\\
17 & $568$ & \texttt{10110000} & 103 & $345$ & \texttt{00011100} & 189 & $14678$ & \texttt{11101001}\\
18 & $58$ & \texttt{10010000} & 104 & $2345$ & \texttt{00011110} & 190 & $1478$ & \texttt{11001001}\\
19 & $578$ & \texttt{11010000} & 105 & $234$ & \texttt{00001110} & 191 & $14578$ & \texttt{11011001}\\
20 & $57$ & \texttt{01010000} & 106 & $2346$ & \texttt{00101110} & 192 & $1458$ & \texttt{10011001}\\
21 & $457$ & \texttt{01011000} & 107 & $236$ & \texttt{00100110} & 193 & $13458$ & \texttt{10011101}\\
22 & $47$ & \texttt{01001000} & 108 & $1236$ & \texttt{00100111} & 194 & $1348$ & \texttt{10001101}\\
23 & $478$ & \texttt{11001000} & 109 & $136$ & \texttt{00100101} & 195 & $12348$ & \texttt{10001111}\\
24 & $48$ & \texttt{10001000} & 110 & $1346$ & \texttt{00101101} & 196 & $1248$ & \texttt{10001011}\\
25 & $468$ & \texttt{10101000} & 111 & $146$ & \texttt{00101001} & 197 & $12478$ & \texttt{11001011}\\
26 & $46$ & \texttt{00101000} & 112 & $1467$ & \texttt{01101001} & 198 & $124678$ & \texttt{11101011}\\
27 & $346$ & \texttt{00101100} & 113 & $147$ & \texttt{01001001} & 199 & $24678$ & \texttt{11101010}\\
28 & $36$ & \texttt{00100100} & 114 & $1457$ & \texttt{01011001} & 200 & $234678$ & \texttt{11101110}\\
29 & $368$ & \texttt{10100100} & 115 & $145$ & \texttt{00011001} & 201 & $23467$ & \texttt{01101110}\\
30 & $38$ & \texttt{10000100} & 116 & $1345$ & \texttt{00011101} & 202 & $234567$ & \texttt{01111110}\\
31 & $378$ & \texttt{11000100} & 117 & $134$ & \texttt{00001101} & 203 & $24567$ & \texttt{01111010}\\
32 & $37$ & \texttt{01000100} & 118 & $1234$ & \texttt{00001111} & 204 & $245678$ & \texttt{11111010}\\
33 & $357$ & \texttt{01010100} & 119 & $123$ & \texttt{00000111} & 205 & $24568$ & \texttt{10111010}\\
34 & $35$ & \texttt{00010100} & 120 & $1238$ & \texttt{10000111} & 206 & $234568$ & \texttt{10111110}\\
35 & $235$ & \texttt{00010110} & 121 & $128$ & \texttt{10000011} & 207 & $23458$ & \texttt{10011110}\\
36 & $25$ & \texttt{00010010} & 122 & $1278$ & \texttt{11000011} & 208 & $234578$ & \texttt{11011110}\\
37 & $258$ & \texttt{10010010} & 123 & $127$ & \texttt{01000011} & 209 & $23578$ & \texttt{11010110}\\
38 & $28$ & \texttt{10000010} & 124 & $1267$ & \texttt{01100011} & 210 & $235678$ & \texttt{11110110}\\
39 & $278$ & \texttt{11000010} & 125 & $126$ & \texttt{00100011} & 211 & $23567$ & \texttt{01110110}\\
40 & $27$ & \texttt{01000010} & 126 & $1256$ & \texttt{00110011} & 212 & $123567$ & \texttt{01110111}\\
41 & $267$ & \texttt{01100010} & 127 & $125$ & \texttt{00010011} & 213 & $12367$ & \texttt{01100111}\\
42 & $26$ & \texttt{00100010} & 128 & $1258$ & \texttt{10010011} & 214 & $123467$ & \texttt{01101111}\\
43 & $246$ & \texttt{00101010} & 129 & $12568$ & \texttt{10110011} & 215 & $12346$ & \texttt{00101111}\\
44 & $24$ & \texttt{00001010} & 130 & $1268$ & \texttt{10100011} & 216 & $123456$ & \texttt{00111111}\\
45 & $124$ & \texttt{00001011} & 131 & $12678$ & \texttt{11100011} & 217 & $12356$ & \texttt{00110111}\\
46 & $14$ & \texttt{00001001} & 132 & $2678$ & \texttt{11100010} & 218 & $123568$ & \texttt{10110111}\\
47 & $148$ & \texttt{10001001} & 133 & $25678$ & \texttt{11110010} & 219 & $12368$ & \texttt{10100111}\\
48 & $18$ & \texttt{10000001} & 134 & $2578$ & \texttt{11010010} & 220 & $123678$ & \texttt{11100111}\\
49 & $178$ & \texttt{11000001} & 135 & $24578$ & \texttt{11011010} & 221 & $13678$ & \texttt{11100101}\\
50 & $17$ & \texttt{01000001} & 136 & $4578$ & \texttt{11011000} & 222 & $135678$ & \texttt{11110101}\\
51 & $167$ & \texttt{01100001} & 137 & $45678$ & \texttt{11111000} & 223 & $15678$ & \texttt{11110001}\\
52 & $16$ & \texttt{00100001} & 138 & $4678$ & \texttt{11101000} & 224 & $145678$ & \texttt{11111001}\\
53 & $156$ & \texttt{00110001} & 139 & $34678$ & \texttt{11101100} & 225 & $14568$ & \texttt{10111001}\\
54 & $15$ & \texttt{00010001} & 140 & $3678$ & \texttt{11100100} & 226 & $124568$ & \texttt{10111011}\\
55 & $135$ & \texttt{00010101} & 141 & $23678$ & \texttt{11100110} & 227 & $12458$ & \texttt{10011011}\\
56 & $13$ & \texttt{00000101} & 142 & $2368$ & \texttt{10100110} & 228 & $124578$ & \texttt{11011011}\\
57 & $138$ & \texttt{10000101} & 143 & $23568$ & \texttt{10110110} & 229 & $12578$ & \texttt{11010011}\\
58 & $1378$ & \texttt{11000101} & 144 & $2356$ & \texttt{00110110} & 230 & $123578$ & \texttt{11010111}\\
59 & $137$ & \texttt{01000101} & 145 & $23456$ & \texttt{00111110} & 231 & $12378$ & \texttt{11000111}\\
60 & $1357$ & \texttt{01010101} & 146 & $2456$ & \texttt{00111010} & 232 & $123478$ & \texttt{11001111}\\
61 & $157$ & \texttt{01010001} & 147 & $12456$ & \texttt{00111011} & 233 & $13478$ & \texttt{11001101}\\
62 & $1578$ & \texttt{11010001} & 148 & $1456$ & \texttt{00111001} & 234 & $134578$ & \texttt{11011101}\\
63 & $158$ & \texttt{10010001} & 149 & $14567$ & \texttt{01111001} & 235 & $34578$ & \texttt{11011100}\\
64 & $1568$ & \texttt{10110001} & 150 & $1567$ & \texttt{01110001} & 236 & $345678$ & \texttt{11111100}\\
65 & $168$ & \texttt{10100001} & 151 & $12567$ & \texttt{01110011} & 237 & $34568$ & \texttt{10111100}\\
66 & $1678$ & \texttt{11100001} & 152 & $1257$ & \texttt{01010011} & 238 & $134568$ & \texttt{10111101}\\
67 & $678$ & \texttt{11100000} & 153 & $12457$ & \texttt{01011011} & 239 & $13456$ & \texttt{00111101}\\
68 & $5678$ & \texttt{11110000} & 154 & $1245$ & \texttt{00011011} & 240 & $134567$ & \texttt{01111101}\\
69 & $567$ & \texttt{01110000} & 155 & $12345$ & \texttt{00011111} & 241 & $1345678$ & \texttt{11111101}\\
70 & $4567$ & \texttt{01111000} & 156 & $1235$ & \texttt{00010111} & 242 & $134678$ & \texttt{11101101}\\
71 & $467$ & \texttt{01101000} & 157 & $12358$ & \texttt{10010111} & 243 & $1234678$ & \texttt{11101111}\\
72 & $2467$ & \texttt{01101010} & 158 & $1358$ & \texttt{10010101} & 244 & $123468$ & \texttt{10101111}\\
73 & $247$ & \texttt{01001010} & 159 & $13578$ & \texttt{11010101} & 245 & $1234568$ & \texttt{10111111}\\
74 & $2478$ & \texttt{11001010} & 160 & $3578$ & \texttt{11010100} & 246 & $123458$ & \texttt{10011111}\\
75 & $248$ & \texttt{10001010} & 161 & $35678$ & \texttt{11110100} & 247 & $1234578$ & \texttt{11011111}\\
76 & $2468$ & \texttt{10101010} & 162 & $3567$ & \texttt{01110100} & 248 & $123457$ & \texttt{01011111}\\
77 & $268$ & \texttt{10100010} & 163 & $34567$ & \texttt{01111100} & 249 & $1234567$ & \texttt{01111111}\\
78 & $2568$ & \texttt{10110010} & 164 & $3457$ & \texttt{01011100} & 250 & $124567$ & \texttt{01111011}\\
79 & $256$ & \texttt{00110010} & 165 & $13457$ & \texttt{01011101} & 251 & $1245678$ & \texttt{11111011}\\
80 & $2567$ & \texttt{01110010} & 166 & $1347$ & \texttt{01001101} & 252 & $125678$ & \texttt{11110011}\\
81 & $257$ & \texttt{01010010} & 167 & $13467$ & \texttt{01101101} & 253 & $1235678$ & \texttt{11110111}\\
82 & $2457$ & \texttt{01011010} & 168 & $1367$ & \texttt{01100101} & 254 & $12345678$ & \texttt{11111111}\\
83 & $245$ & \texttt{00011010} & 169 & $13567$ & \texttt{01110101} & 255 & $2345678$ & \texttt{11111110}\\
84 & $2458$ & \texttt{10011010} & 170 & $1356$ & \texttt{00110101} &  &  & \\
85 & $458$ & \texttt{10011000} & 171 & $13568$ & \texttt{10110101} &  &  & \\
\bottomrule
\end{longtable}
\endgroup

\bibliographystyle{unsrtnat}
\bibliography{refs}

\end{document}